\documentclass[fleqn,10pt]{wlscirep}
\usepackage[utf8]{inputenc}
\usepackage[T1]{fontenc}
\usepackage{subcaption}
\usepackage{float}
\usepackage{pdfpages}
\usepackage{setspace}
\usepackage{graphicx}
\usepackage{adjustbox}
\usepackage{tabularx}
\usepackage{changepage}
\usepackage[table]{xcolor}
\usepackage{colortbl}
\usepackage{booktabs}
\usepackage{pgf}
\usepackage[left]{lineno}
\usepackage{caption}

\title{Disentangled Fingerprints suggest no historical weakening of Atlantic Overturning and Subpolar Gyre}

\author[1*]{Bahar Emirzade}
\author[2]{Jade Ajagun-Brauns}
\author[1,3]{Maya Ben-Yami}
\author[1,3]{Sebastian Bathiany}
\author[4]{Yechul Shin}
\author[4,5]{Jong-Seong Kug}
\author[1,3,6]{Niklas Boers}
\affil[1]{Earth System Modelling, School of Engineering and Design, Technical University Munich, Munich, Germany}
\affil[2]{University of Copenhagen, Niels Bohr Institute, Physics of Ice, Climate and Earth, København, Denmark}
\affil[3]{Potsdam Institute for Climate Impact Research, Potsdam, Germany}
\affil[4]{School of Earth and Environmental Sciences, Seoul National University, Seoul, Republic of Korea}
\affil[5]{Institute for Sustainable Development, Seoul National University, Republic of Korea}
\affil[6]{Global Systems Institute and Department of Mathematics, University of Exeter, Exeter, UK}
\affil[*]{bahar.emirzade@tum.de}

\begin{abstract}

The Atlantic Meridional Overturning Circulation (AMOC) and the Subpolar Gyre (SPG) are key components of the Earth’s climate system, both potentially prone to destabilization and abrupt shifts under anthropogenic forcing. Given the strong influence of the AMOC and the SPG on global climate, detecting changes in their stability is of high importance, yet is hindered by the limited length of observational records. In the absence of long-term direct AMOC observations, several observation-based fingerprints of the AMOC have been proposed. However, existing fingerprints do not clearly distinguish between SPG and AMOC variability, instead reflecting an opaque mixture of dynamics from both systems.

Here, we assess the performance of widely used AMOC fingerprints across Coupled Model Intercomparison Project 6 (CMIP6) models and compare them with statistically optimal fingerprints that we derive from sea surface temperature and salinity. We show that traditional AMOC fingerprints exhibit weak correlations with both AMOC and SPG strength. In contrast, our statistically optimal fingerprints, trained on pre-industrial control (piControl) simulations, consistently outperform existing approaches across independent segments of piControl simulations and historical experiments. These fingerprints accurately reconstruct AMOC and SPG variability while minimizing overlap, enabling a clear separation between the two systems.

Our results provide the first tailored observational fingerprints that robustly distinguish AMOC and SPG, demonstrating the potential of statistically optimized approaches to improve detection and attribution of changes in Atlantic circulation, including precursor signals of potential tipping behavior. Applying our approach to sea surface temperature observations, we find that our optimized fingerprints do not confirm the substantial AMOC weakening in the recent decades as suggested by traditional fingerprints, but also do not rule out a loss of stability.

\end{abstract}
\begin{document}

\flushbottom
\maketitle

\thispagestyle{empty}

\section*{Introduction}


The Atlantic Meridional Overturning Circulation (AMOC) is a large-scale ocean circulation system resulting from a combination of thermohaline and wind-driven processes \cite{buckley2016, rahmstorf2024}. Warm surface waters from the tropics are transported to the North Atlantic (NA) by the AMOC, where they cool down, consequently gaining density, and sink to form the North Atlantic Deep Water (NADW) \cite{rahmstorf2024}. This forms the deep water in regions such as the Greenland, Iceland and the Norwegian (GIN) Seas\cite{petit2020}. The amount of deep water formed in these regions influences the strength of the AMOC, which is maintained at its current levels through the positive salt-advection feedback. By this feedback mechanism, a stronger AMOC transports more saline water northwards towards deep water formation sites in the NA, leading to denser water masses, more sinking, and thus a stronger AMOC\cite{weijer2019}. In contrast, loss of density of the water masses due to warming and freshening at the relevant sites can lead to a weaker AMOC and, potentially, to an abrupt shift toward a considerably weaker circulation state \cite{boers2021,rahmstorf2024,vanwesten2024}. 

Paleoclimate records\cite{henry2016,boers2018a,boers2018b,weijer2019,liu2023,rahmstorf2024}, but also models ranging from simple ones \cite{stommel1961 ,manabe1988} to complex Earth System Models (ESM)\cite{liu2017,weijer2019,jackson2023,vanwesten2024} provide evidence that the AMOC is a bistable system \cite{ipcc2013}. The bistability has been postulated such that sufficient external forcing - like increased freshwater input near deep water formation regions - could abruptly shift the AMOC from the current, strong circulation mode to a substantially weaker stable state\cite{ipcc2021} by decreasing the formation of deep water. During this transition, the system would gradually lose stability until it reaches the critical point, the crossing of which would trigger a collapse into the weaker state. It would remain in this new, weak state, unless the forcing is sufficiently reduced below the level that sustained the previous strong state.

Increasing global temperatures are expected to enhance freshwater input into the NA through an intensified hydrological cycle and Greenland Ice Sheet melt, contributing to surface freshening. There is evidence suggesting that the AMOC has weakened over the past century, potentially linked to this surface freshening \cite{rahmstorf2015, zhu2023, caesar2021}. Additionally, observation-based statistical stability indicators provide signs that the AMOC may be approaching an abrupt transition, with potentially major climatic, ecological, and socioeconomic impacts \cite{boers2021,benyami2023}. However, assessing stability in such a slow component of the climate system requires long time series, while direct measurements of AMOC strength span only about two decades (RAPID array since 2004) \cite{johns2023}. Consequently, longer-term changes in the AMOC are often inferred from indirect observations, including proxy records and physically motivated fingerprints.

Commonly, AMOC fingerprints are based on observable variables in the ocean which have a physically understood link to AMOC strength. The most prevalent of these fingerprints are based on sea surface temperature (SST) \cite{caesar2018, jackson2020, rahmstorf2015} and sea surface salinity (SSS) averages\cite{zhu2023} over different regions, and fingerprints based on sea-surface heat fluxes have also been proposed\cite{terhaar2025}. A widely used fingerprint for the AMOC is given by the SSTs averaged over the subpolar gyre (SPG) region south of Greenland, relative to the northern hemisphere or global SST mean\cite{caesar2018,rahmstorf2015}. This fingerprint, also known as the “warming hole” fingerprint, is based on the understanding that a weaker AMOC transports less heat to the NA, leading to relatively cooler surface waters\cite{rahmstorf2024}. In addition to a reduction in the mean circulation strength\cite{caesar2018,rahmstorf2015}, statistical analyses of this and other SST- and SSS-based fingerprints suggest that the stability of the AMOC has declined in the course of the last century \cite{boers2021,benyami2023,benyami2024, ditlevsen2023}. 


While SST averages in and around the North Atlantic have been widely used as indicators of AMOC variability, these surface temperature patterns are not solely governed by the overturning circulation. The SPG as a gyre plays a key role in redistributing heat, salt, and nutrients across the basin \cite{Hatun:2005}, and, as a comparatively fast ocean component, influences the regional climate on interannual to decadal timescales\cite{Fan2023}. The SPG is driven by both wind stress and buoyancy forcing. Positive wind stress curl over the subpolar North Atlantic induces Ekman divergence, leading to upwelling and a lowering of the sea surface height that supports the cyclonic circulation\cite{MarshallSchott1999}.

Buoyancy forcing, in turn, controls the density structure of the gyre and thereby modulates its circulation strength \cite{Montoya2011}. The boundary regions are influenced by the inflow of warm, saline subtropical waters via the North Atlantic Current and colder, fresher polar waters via the Labrador and East Greenland Currents. Wintertime cooling over the surface of the gyre interior increases surface density and drives deep convection in the Labrador and Irminger Seas\cite{Spall2004, petit2020}. This combination establishes horizontal density gradients between a dense interior and lighter boundary waters. Consequently, pressure gradients are generated which, via geostrophic balance and thermal wind shear, sustain a predominantly baroclinic circulation\cite{Montoya2011}.

Critically, such density gradients enable nonlinear feedbacks between circulation strength, salinity transport, and deep convection, and lead to the suspected bistable behavior of the SPG \cite{born2014}. A stronger gyre enhances the retention and production of dense waters in the interior, reinforcing density gradients and sustaining deep convection, whereas a weakened gyre allows increased inflow of warmer and fresher subtropical waters, reducing density, suppressing convection, and further weakening the circulation. This advective-convective feedback provides a physical basis for the bistability of the SPG \cite{born2014}, whereby the system can reside in either a strong, convective state or a weak, stratified state.

Recent work by Falkena et al. (2025) \cite{falkena2024} refines this picture by demonstrating that the feedbacks governing SPG variability are strongly timescale-dependent. Increased SSS enhances deep convection in the interior of the SPG, which cools and densifies the water column, strengthening the baroclinic flow. On short timescales, a stronger SPG leads to a reduction in SSS due to the increased convection, which feeds back negatively to SPG strength, whereas on longer timescales the impact of a strengthened SPG on salinity is of opposite sign \cite{falkena2024}. This highlights that the stability of the SPG emerges from the interplay of competing feedbacks acting on different timescales.

The coexistence of positive and negative feedback mechanisms has led the SPG to be identified as a potential climate tipping element in its own right \cite{born2016,swingedouw2021,Armstrong2022,arellano2022}, motivating its investigation as a system distinct from the AMOC. One of the open questions is how an SPG collapse would influence the AMOC; it has been shown in simulations that an SPG collapse under future warming may not necessarily imply a concurrent collapse of the AMOC \cite{sgubin2017}. It should also be noted that either system can show a weakening without destabilisation \cite{weijer2019, jackson2023}.

The SPG, nevertheless, influences the AMOC through its impact on deep-water formation \cite{born2013}. However, there is currently no consensus on the precise nature of the SPG-AMOC relationship \cite{klockmann2020}, nor on how this coupling may evolve under anthropogenic climate change \cite{zhu2024}. One proposed mechanism is that a strong SPG enhances surface salinity and density in North Atlantic deep-water formation regions, thereby promoting stronger deep convection and a more vigorous AMOC \cite{latif2022,born2013}.

As the SPG and the AMOC may weaken and undergo tipping independently, fingerprints should be able to distinguish between the two systems. Ideally, an optimal fingerprint would exhibit a high correlation only with the system of interest \cite{jackson2020}. Furthermore, fingerprints should be designed with their specific application in mind. If we are interested in the overall trend, the trend of the fingerprint needs to be related to the trend of the system of interest \cite{caesar2018,Menary2020}. If we are interested in whether the system is losing stability, e.g. measured in terms of critical slowing down, we require a fingerprint that captures the variability of the system \cite{boers2021,boers2025}.

Due to the aforementioned lack of direct long-term AMOC observations, and a similar lack of direct SPG observations \cite{ghosh2023}, to understand the relationship of the AMOC and SPG to their potential fingerprints, we use state-of-the-art ESMs from the Coupled Model Intercomparison Project Phase 6 (CMIP6) \cite{jackson2020}. The advantage of using models is that we can analyse the systems over longer periods of time, accounting for variability, and investigate different forcing scenarios using different experiments. For example, in CMIP6 we utilise the pre-industrial control run (piControl), where there is no anthropogenic forcing, and the historical run, which represents the anthropogenic forcing between the years 1850 to 2014. 

In this study, we analyse traditional, SST-based AMOC fingerprints in pre-industrial and historical CMIP6 model simulations, assessing their ability to reconstruct the AMOC timeseries. Furthermore, we analyse to what extent the signal captured by these traditional fingerprints is due to the AMOC instead of the SPG. We construct two sets of statistically optimal fingerprints using Ridge regression (RR), one for the AMOC and one for the SPG. Their capacity to capture variability in their respective systems is computed for the piControl and historical simulations in the same CMIP6 models. Finally, we compare their performance to the traditional fingerprints, and discuss the correlation between the AMOC and the SPG, and the difference of this correlation between the piControl and the historical simulations.

\newpage
\section*{Results}

\begin{figure}[h!]
  \centering
  \includegraphics[width=0.9\textwidth]{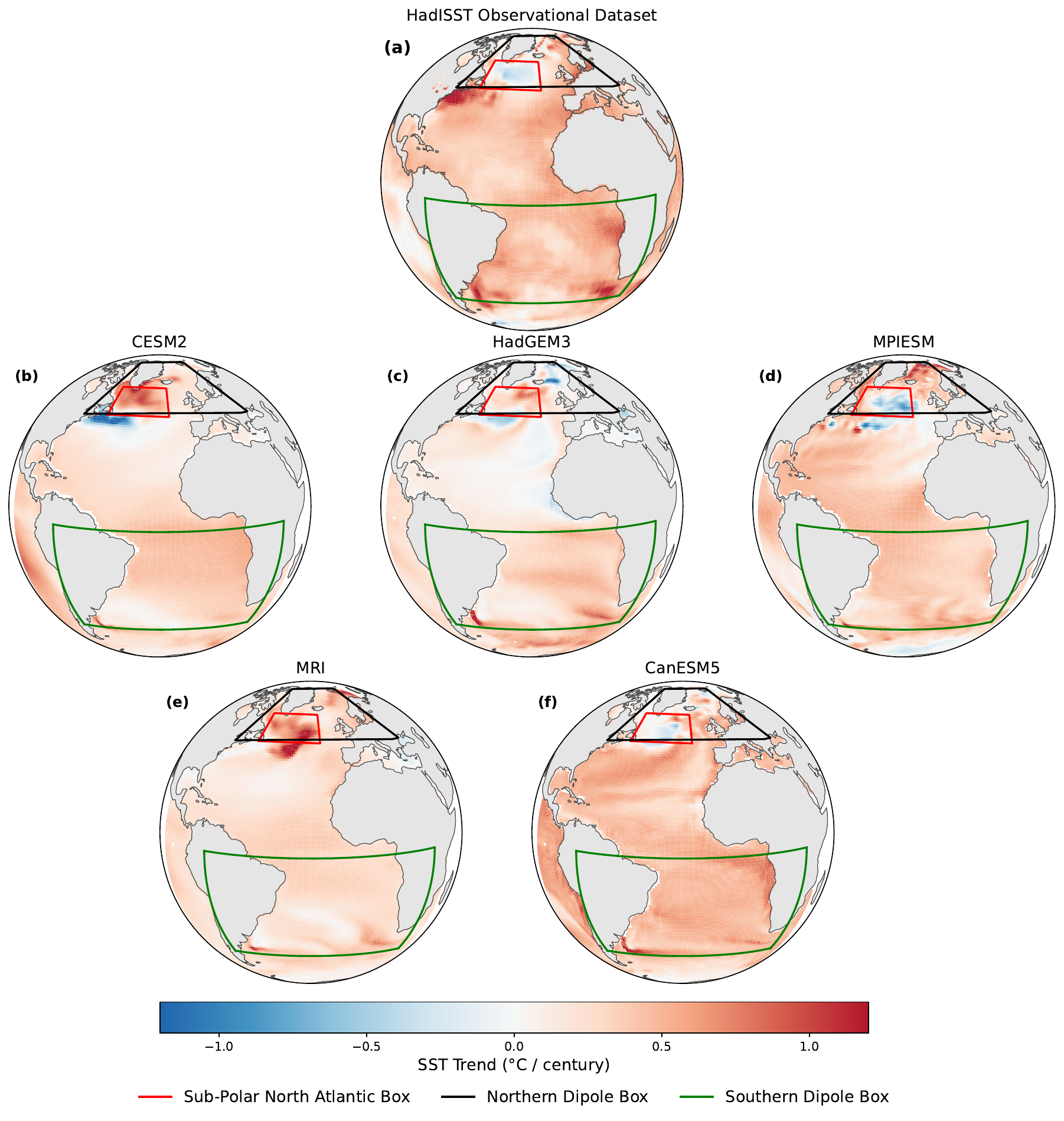}
  \caption{\textbf{Linear trend of Sea Surface Temperatures for HadISST observations and historical CMIP6 simulations.} The map shows the linear 1850-2024 trend in the HadISST dataset and similarly for the historical timeseries of each CMIP6 model used. 
 The SSTs are normalized by subtracting the global, all-time mean, and then the trend is calculated using a linear regression. A region with negative SST trend, the warming hole, can be observed south of Greenland for the HadISST observational dataset. The regions used to calculate the traditional AMOC fingerprint regions are outlined in red for the Sub-Polar North Atlantic box, green for the Southern Dipole box and black for the Northern Dipole box (see Methods). The warming hole is not present for the MRI, HadGEM and CESM models. Furthermore, for MPIESM and CanESM, the location and extent of the warming hole is varied compared to HadISST.
 }
  \label{fig:SST_trends}
\end{figure}

For a selection of five CMIP6 models (see Methods for details), we compute three commonly used, traditional SST-based AMOC fingerprints referred to as $\mathrm{SST_{SG-G}}$, $\mathrm{SST_{SG-NH}}$ and $\mathrm{SST_{DP}}$, (see Methods) and compare them to the corresponding AMOC and SPG timeseries from the model outputs, referred to as the "true AMOC" and "true SPG".  To evaluate how well each fingerprint reflects these ocean systems, we calculate Pearson correlation coefficients and Mean Squared Errors (MSE) between the true system and the fingerprint timeseries. Furthermore, to reveal and discuss potential stability loss, we compute the restoring rates of the model fingerprint timeseries, as well as the restoring rates of the observed fingerprints based on the Met Office Hadley Centre Sea Ice and Sea Surface Temperature data set (HadISST)\cite{Rayner2003}, following the procedure outlined by Boers (2021) \cite{boers2021} (see Methods). The restoring rate computations on CMIP6 models help us evaluate the correlation between the true system's restoring rate and that of the fingerprint. Together, these metrics help assess four key properties of each fingerprint: (i) the capacity to capture high frequency variability; (ii) the effectiveness in reproducing long-term trends using historical timeseries; (iii) the ability to distinguish between the SPG and AMOC signals; and (iv) how well the fingerprint signals any stability changes of the system. 

Figure 1 shows the linear trend per century in the HadISST global SST dataset, as well as the historical timeseries of the selected CMIP6 models. The "warming hole" phenomenon is clearly visible in the HadISST dataset, along with an overall warming trend of the global oceans. CMIP6 models do not show the warming hole consistently. Contrastingly, a slight warming trend of the relevant location in CESM, HadGEM and MRI is present. We also note that the location of the warming hole varies between the HadISST dataset and the two models that have this cooling trend.


Utilizing SST and SSS data from CMIP6 models over the entire Atlantic basin, we use ridge regression (RR) \cite{hoerl1970} to construct two separate sets of fingerprints for each of the selected CMIP6 models; one for the AMOC and the other for the SPG. These statistical fingerprints make use of the maximum available information about these systems present in the SST and SSS fields. Our statistically optimal RR fingerprints take as input SST and SSS data that we smooth to varying degrees beforehand, in order to examine the effect of different time scales. Specifically, SST and SSS data are smoothed via rolling mean filters at annual (RR1), 5 year  (RR5), 15 year (RR15), 20 year (RR20) and 30 year (RR30) time scales; the same temporal smoothing is applied to the AMOC $\mathrm{RR_{AMOC}}$ and SPG $\mathrm{RR_{SPG}}$ streamfunctions prior to estimating the regression models.

We train the RR models to learn the mapping from SST and SSS data to the true AMOC and SPG strength (Methods) using a portion of the piControl data, validate it by reconstructing the "unseen" (last 20\,\%) portion of piControl, and test it on the historical simulation. Note that since these fingerprints are trained on each model individually, they are optimized for the corresponding SST and SSS variability patterns of each individual model. 

The RR successfully reconstructs both the unseen part of the piControl and the historical AMOC simulations. Figure 2 shows the RR reconstruction of the 5-year rolling smoothed AMOC ($\mathrm{RR5_{AMOC}}$), and SPG $\mathrm{RR5_{SPG}}$, for the entire historical runs of each model. The RR5 reconstruction has a high correlation with the true time series, above 0.68 for all models in the historical AMOC timeseries.   

\begin{figure}[H]
  \centering
  \includegraphics[width=1\textwidth]{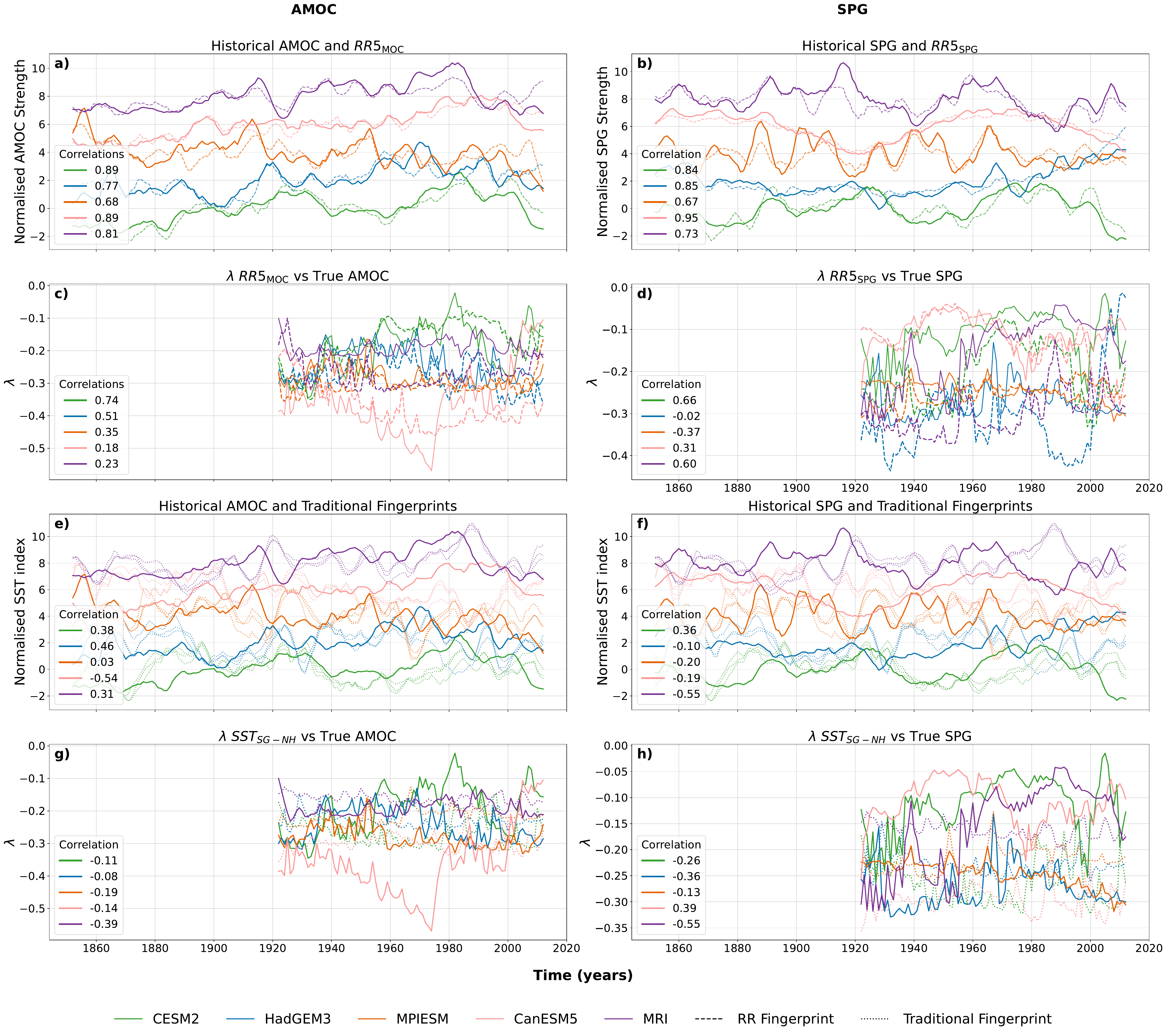}
  
  \caption{\textbf{Modelled historical AMOC (left) and SPG (right) timeseries along with their RR and traditional fingerprint reconstructions and the corresponding restoring rate ($\lambda$) calculations.} All AMOC and SPG strength timeseries, as well as the traditional fingerprints, are smoothed with a 5 year rolling mean (the  corresponding 1-, 15-, 20-,  and 30-year smoothed timeseries can be found in the SI). The timeseries in panels (a), (b), (e) and (f) are offset on the y-axis to prevent overlapping timeseries and improve visibility. Panel (a) shows the true modelled AMOC strength time series (solid lines) and the corresponding RR5 reconstructions (dashed lines) for each ESM, together with the corresponding correlations ranging from 0.89 in CESM and CanESM to 0.68 in MRIESM.  Panel (c) shows the restoring rate calculations with a 70 year window of the detrended timeseries seen on panel (a). The correlation between the restoring rates of the true modelled AMOC strength and the RR5 reconstructions are also stated, with the highest correlation in CESM (0.74) and lowest in CanESM (0.18). No clear positive trend in restoring rate is present for any of the models. Panel (e) shows the same as panel (a) but for traditional fingerprints, showing that across ESMs, the true modelled AMOC is captured much better by RR5 than by the traditional fingerprints. Panel (g) shows the restoring rates of the traditional fingerprints shown in (e); demonstrating that the restoring rates of the true modelled AMOC are also better captured by our RR5 fingerprints than by the traditional ones. The correlation values shown refer to the $\mathrm{SST_{SG-G}}$ fingerprint as proposed by \cite{caesar2018}.  Panels (b), (d), (f) and (h) show the same information but for the SPG timeseries. Note that the traditional fingerprint timeseries and, hence, their restoring rates are the same for panels (e) and (f) and (g) and (h) since traditional fingerprints aim to capture only AMOC streamfunction and not the SPG. While the RR5 reconstruction of SPG is as good as for AMOC, the restoring rate correlation is better for some models and worse for others. Panel (h) shows that traditional fingerprints do not have high skill in reconstructing the SPG timeseries with correlations ranging from 0.36 in CESM to -0.55 in MRI.}
  \label{fig:historical_ts}
\end{figure}

Compared to the RR fingerprint for the AMOC, the SPG fingerprint performs similar in the historical timeseries, with correlations ranging from 0.67 in MPIESM to 0.95 in CanESM. We note that MRI has a shorter piControl timeseries, giving RR less data overall to train on, which could be the explanation for the lower performance in MRI compared to other ESMs. A possible reason for the more varied and lower correlation of the SPG fingerprint among models is that the spatial scale of SPG is much smaller compared to that of AMOC, which could imply that the area where the SST and SSS patterns are relevant for the SPG is smaller than for the AMOC. 

Figure 2 also includes the restoring rates for the true modelled historical AMOC and SPG time series, and the corresponding fingerprint reconstructions in the historical timeseries. In line with recent literature \cite{benyami2026} there is no increasing trend in the restoring rates in three out of five models, with CanESM and MPIESM being the models with rising restoring rate trend and HadGEM instead showing a decrease in restoring rate. Compared to traditional fingerprints, the restoring rates calculated from our RR fingerprint exhibit substantially higher correlations to the restoring rate of the true AMOC and SPG timeseries. In fact, our results show that the traditional fingerprints cannot capture the historical restoring rates of neither AMOC nor SPG. 

In Supplementary Information Figure 1 (SI1), we show the multi-model average of the RR coefficient distribution over the Atlantic ocean for the AMOC RR5 fingerprint and the SPG RR5 fingerprint. The coefficient projections are also separated for SSS and SST. The coefficients represent the extent to which RR uses these locations for the reconstruction. The RR coefficients indicate  negative relationship between the SPG strength and the SSTs over the Labrador basin, south of Greenland, whereas the opposite is true for the RR AMOC. Lower SSTs in this region indicate a stronger SPG in the shorter term, as the colder waters imply a denser interior\cite{falkena2024}.  It is also physically understood that a stronger AMOC would bring warmer surface waters to the North Atlantic\cite{rahmstorf2015}, explaining how the SSTs in the same region would have a positive relationship to AMOC strength.  

In Supplementary Figure 2 (SI2), we show the RR weight distributions across the Atlantic ocean for each RR5 fingerprint. We note that for the models in which the $\mathrm{RR5_{AMOC}}$ fingerprint reconstruction performs better, RR5 places high coefficients for SSTs in the 30-60\,°\,N to 20-40\,°\,W box. The corresponding weight distribution is the most scattered for MPIESM, which has the lowest performance for the $\mathrm{RR5_{AMOC}}$ fingerprint in the historical simulations.

Comparing the performance of the traditional fingerprints with the $\mathrm{RR5_{AMOC}}$ fingerprint, we note that both the MSE value and the correlation to the true values are significantly improved with the $\mathrm{RR5_{AMOC}}$ fingerprint. Since traditional SST-based fingerprints are designed for capturing the overall trends in AMOC \cite{caesar2018}, it is not surprising that the $\mathrm{RR5_{AMOC}}$ fingerprints perform better in the piControl run, where there is no trend. However, our RR-based optimal fingerprints outperform traditional fingerprints even when we smooth the historical AMOC timeseries and the fingerprint reconstruction with a 15-, 20- or 30-year rolling mean to remove high-frequency variability. Figure 3 illustrates the performance of the traditional and the $\mathrm{RR5_{AMOC}}$ fingerprints on CESM AMOC timeseries. Corresponding information on other models can be seen in Supplementary Table 1. The top panel of Figure 3 shows the entire piControl run and the performance of all fingerprints on this timeseries. As before, the RR fingerprint is used to reconstruct the last 20\,\% of this timeseries. It can be seen that the traditional fingerprints perform similarly to one another. While having correlations around 0.5 in the piControl run, the correlations of the traditional SST based fingerprints decrease in the historical run. The traditional fingerprints reconstruct the historical AMOC strength better in the latter half of the timeseries, compared to the first half.

\begin{figure}[H]
\centering
\includegraphics[width=0.9\textwidth]{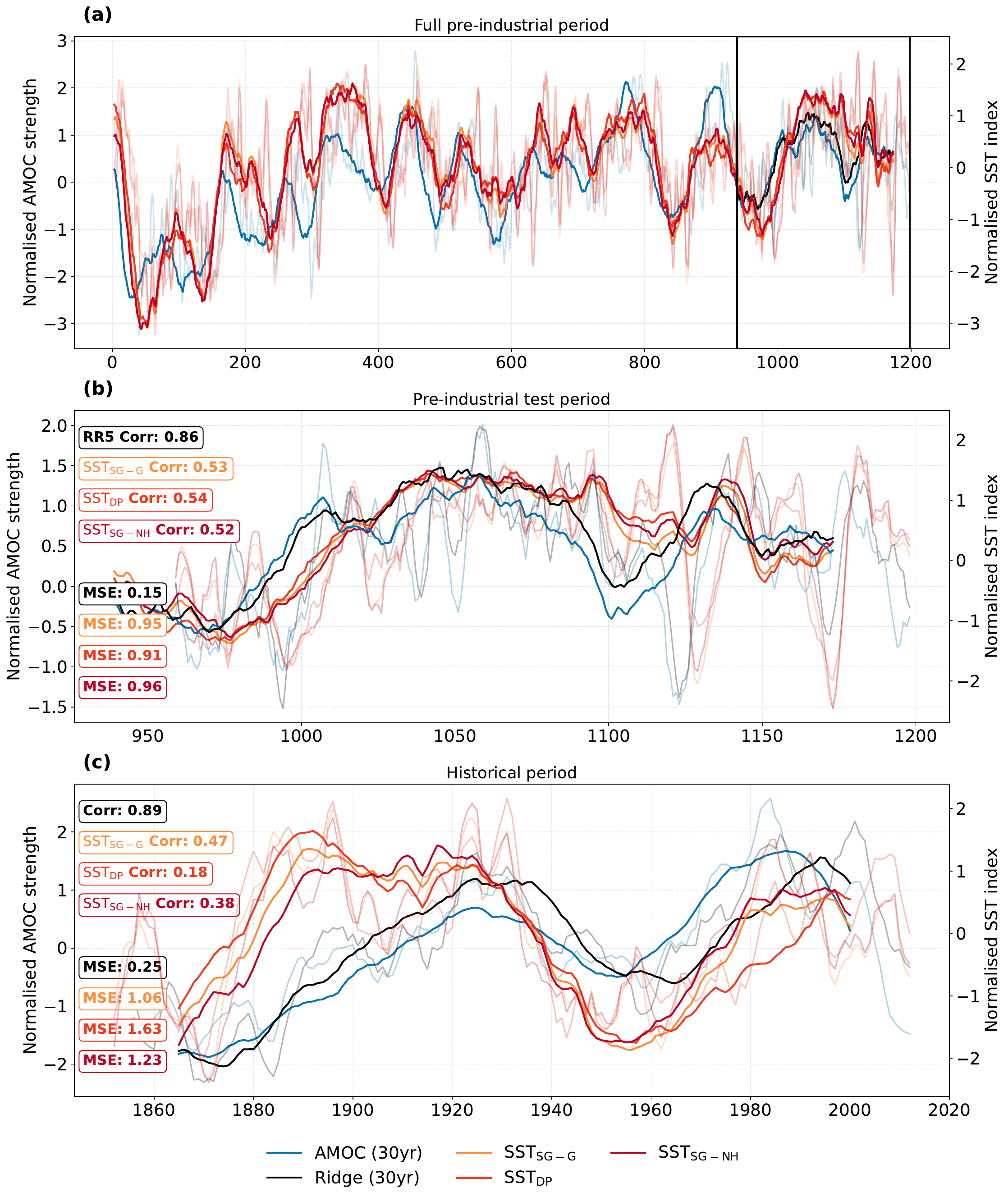}
\caption{\textbf{Comparison of AMOC fingerprints in CESM piControl and historical simulations, shown with 30-year rolling means.} Panel (a) showcases the entire piControl timeseries for CESM, where the blue is the true AMOC, black is the RR30 fingerprint, red is the  $\mathrm{SST_{DP}}$ fingerprint, yellow is the  $\mathrm{SST_{SG-NH}}$ fingerprint and orange is the  $\mathrm{SST_{SG-G}}$ fingerprint. Since the RR fingerprint is used to reconstruct only the last 20\,\% of the timeseries, panel (b) is provided for an easy comparison of all fingerprints in this portion of the piControl timeseries. Panel (c) is the historical run for CESM. For each fingerprint and the true AMOC the 5-year rolling mean values and RR5 fingerprint are plotted with lower opacity and the bold lines represent the corresponding 30-year rolling means and RR30. Note that the RR and true AMOC reconstructions are plotted on a different y-axis from the SST fingerprints. Also note the slight (5-6-year) lag of traditional fingerprints compared to the true AMOC, which is absent for the RR fingerprint.}
\label{fig:CESM2_comparison}
\end{figure}

The performance of the traditional fingerprints is similar in all other models, overall performing better in piControl than in the historical simulation for the 5-year rolling mean AMOC (see SI Table 1). While the traditional fingerprints are better at reproducing the 30-year rolling mean in the piControl AMOC than the 5-year rolling mean, this is not the case for the historical run. Considering that there is no background forcing in the piControl run, we can deduce that the traditional fingerprints do not perform well in capturing variability in the AMOC in piControl or in the historical runs for the models tested. In addition, RR outperforms the traditional fingerprints in capturing trends in the historical timeseries.

Regarding the performance of traditional fingerprints on the SPG timeseries, there are slight negative to no correlations in all models for the  piControl run, yet as much of a positive correlation exists in the SPG timeseries historical runs as is present for the historical AMOC timeseries for CESM and MPIESM (See SI Table 1).

Analysing the correlation between the AMOC and the SPG in the chosen CMIP6 models, we notice that the two systems are mostly positively correlated in both the piControl and historical runs. Figure 4 shows this relationship for each model, and additionally shows whether this relationship is well-represented by the RR fingerprint. The results showcase that the correlation between the AMOC and SPG is model-dependent, being particularly highly correlated in CESM and much less so in MRI and CanESM. For all models but CESM and MRI, the SPG and AMOC have higher correlation in piControl than in the historical runs. It is also worth noting that most correlations lie close to the y=x axis in Figure 4, showing that the RR SPG and AMOC are good at capturing the change in relationship between the SPG and AMOC. This confirms that the RR fingerprints not only have a high correlation with the true time series of both the SPG and AMOC, but are also able to reproduce the correlation between them. 

\begin{figure}[H]
\centering
\begin{minipage}{1\textwidth}
    \centering
    \includegraphics[width=\textwidth]{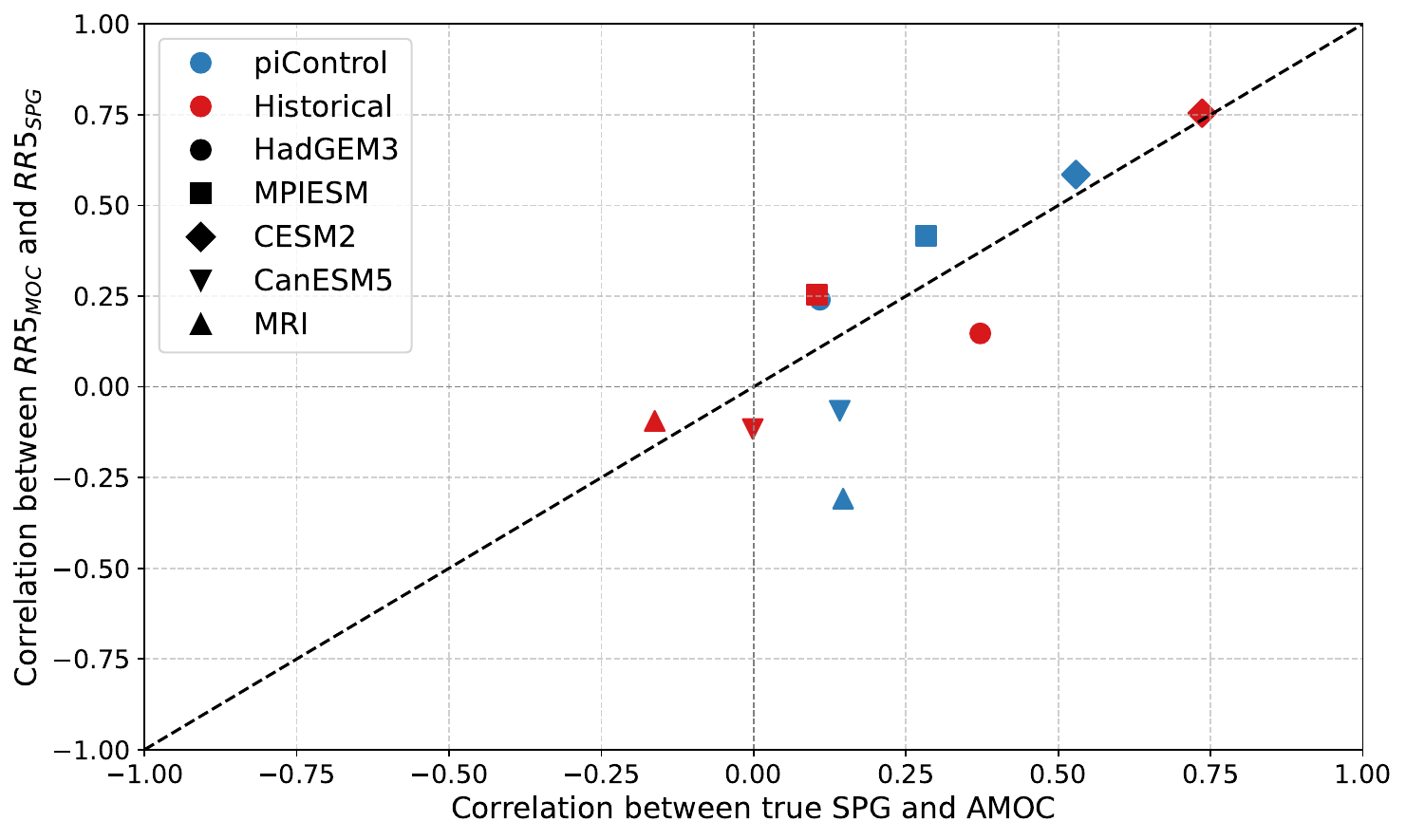}
    \caption{\textbf{Pearson correlation coefficients between true AMOC and SPG vs. correlations of corresponding fingerprints in the piControl and the historical simulation in all models.} Correlations are computed on a 5-year rolling mean smoothing and RR5 values. The x-axis shows the correlation between the true SPG and AMOC from the model outputs and the y-axis shows the correlation between the RR5 reconstruction of the SPG and the AMOC. All blue coloured symbols represent the piControl run and red coloured symbols represent the historical run. The models are represented by circle, square, diamond, reverse triangle and triangle symbols respectively for HadGEM, MPI, CESM, CanESM and MRI. Most datapoints are close to the y=x line, showing that the RR reconstruction captures the full range of the relationship between the AMOC and the SPG. Correlations are also mostly in the positive top right quarter, showing that the relationship between the AMOC and the SPG is mostly positive in both piControl and in the historical run. 
}
    \label{fig:SPG_AMOC_relationship}
\end{minipage}

\begin{minipage}{0.45\textwidth}
    
\end{minipage}
\end{figure}

\begin{figure}[H]
\centering
\begin{minipage}{1\textwidth}
    \centering
    \includegraphics[width=\textwidth]{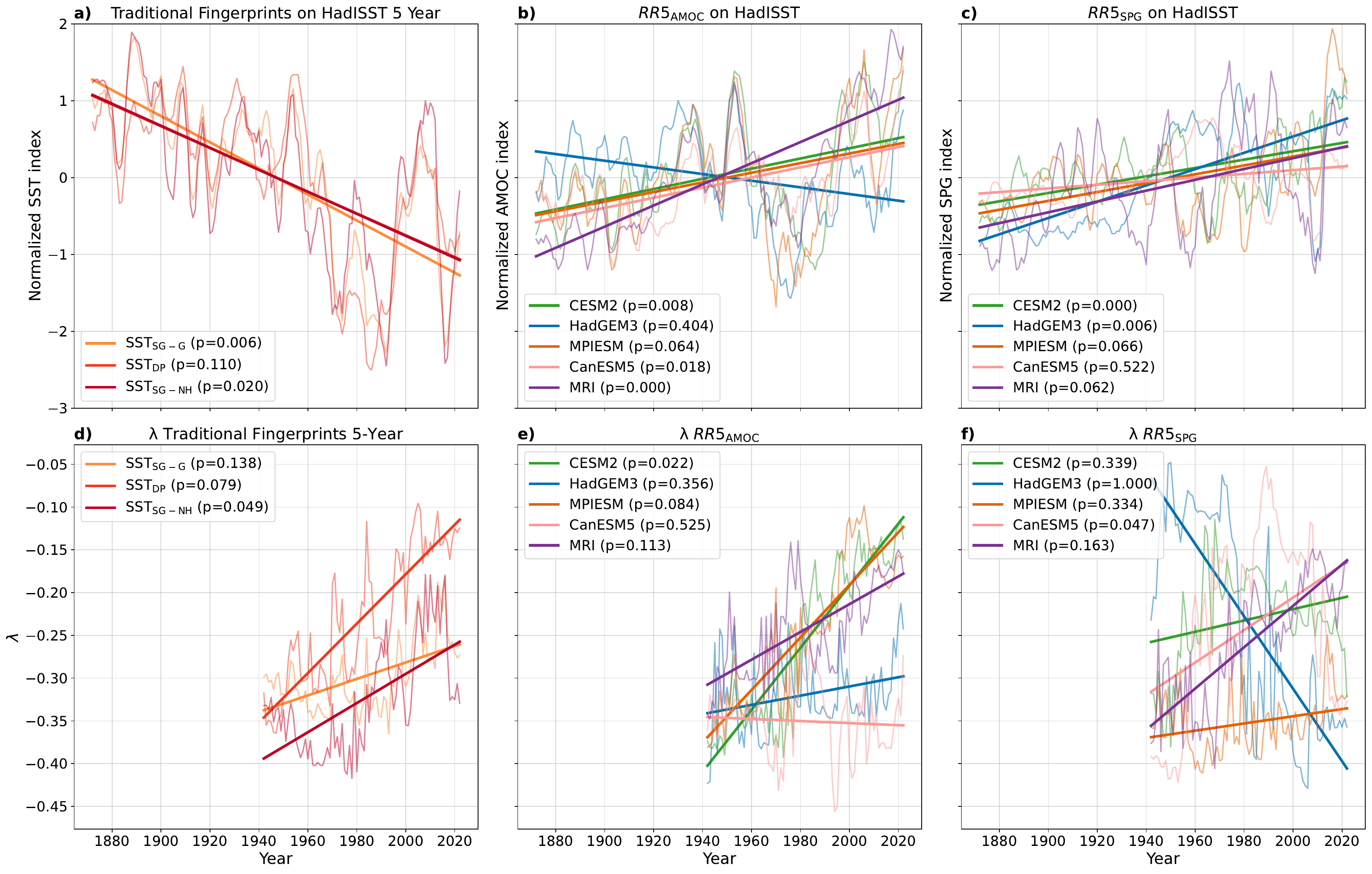}
    \caption{\textbf{HadISST-based reconstructions of AMOC and SPG and their corresponding restoring rate timeseries.}  
    (a) Traditional SST-based fingerprints $\mathrm{SST_{SG-G}}$,  $\mathrm{SST_{DP}}$ and $\mathrm{SST_{SG-NH}}$ all show declining trends from 1880 to 2020, significant for $\mathrm{SST_{SG-G}}$ (p=0.006) and $\mathrm{SST_{SG-NH}}$ (p=0.020) but not for $\mathrm{SST_{DP}}$ (p=0.110). (b) $\mathrm{RR5_{AMOC}}$: RR5 AMOC fingerprints from all 5 CMIP6 models trained only on SST data, taking the normalised and 5-year smoothed HadISST data as input. Reconstructed trends disagree in sign and significance across models, CESM (p=0.008), CanESM (p=0.018) and MRI (p<0.001) show significantly strengthening AMOC, MPIESM shows a non-significant increase, and HadGEM shows a non-significant weakening (p=0.404). (c) As in (b) but for $\mathrm{RR5_{SPG}}$; most models agree on a significant strengthening trend except CanESM (p=0.522). (d-f) restoring rate $\lambda$ for the timeseries seen above each panel, estimated with a 70-year rolling window. While the traditional fingerprints agree on an increasing trend of the restoring rate (panel d), four out of five models  show increasing trends in the restoring rate for $\mathrm{RR5_{AMOC}}$ and $\mathrm{RR5_{SPG}}$ respectively. The corresponding 1 year-, 15-, 20- and 30-year versions of this figure can be found in SI9-12.}
    \label{fig:HadISST_restoring}
\end{minipage}%

\end{figure}

Given the good performance of the RR fingerprint on individual models, we also test their performance on the HadISST dataset and compare with the performance of the traditional fingerprints. We highlight that the RR fingerprints are not designed to be generalisable across models or applied to observations. The goal of applying our RR fingerprints to HadISST is to observe the performance compared to traditional fingerprints, given the  higher performance of RR within models to the systems of interest. Figure 5 shows the different HadISST-based fingerprint timeseries where for panel (a) the traditional fingerprints, panel (b) $\mathrm{RR5_{AMOC}}$ and for panel (c) $\mathrm{RR5_{SPG}}$ have been used to reconstruct the respective systems.  

It can be seen that traditional fingerprints suggest a decrease in strength of the AMOC over the entire HadISST measurement period with two out of three being statistically significant.  When it comes to the $\mathrm{RR5_{AMOC}}$ reconstructions based on the HadISST data, four out of five models show increasing trends, while one shows a non-significant decreasing trend (HadGEM). As for the HadISST based reconstruction of the SPG, the RR5 fingerprints also demonstrate a slight strengthening with four out of five significant. These results are consistent for annual data as well (SI). 

The traditional SST fingerprints agree closely in suggesting a weakening AMOC since 1880, while the model-trained RR5 reconstructions disagree substantially with each other, including disagreement in the sign of the trend (e.g. CESM, MRI, CanESM and MRI strengthening, HadGEM weakening). This divergence highlights how sensitive the SST-to-AMOC / SPG mapping is to the specific ESM used for training, and cautions against treating any single model's reconstruction as ground truth for the observational record. Second, the restoring rate analysis in panels (d) to (f) shows a broadly consistent positive trend, indicating declining stability across the traditional fingerprints and most RR5 reconstructions.

Finally, to compare the ability of the fingerprints to capture and distinguish between variability in the AMOC versus in the SPG, we plot the 5-year rolling mean correlations with respect to each system. Figure 6 shows these correlations, as well as the corresponding MSE scores as radar plots. 

Figure 6 (a) shows that in the piControl run, the RR reconstructions are specific to the system that they are meant to represent. They show high correlation with the true system they are trained to capture, and low correlation with the other circulation system. Consequently, the scatter clouds in correlation space shown in Figure 6 (a) do not overlap, allowing for the two systems to be distinguished via the fingerprints. In contrast, traditional fingerprints do not show large correlations with either system and mostly lie around 0 to 0.5 for AMOC for the piControl runs. In the historical runs, there is a shift of the traditional fingerprints to a negative correlation with the SPG and with the AMOC. However, both in the piControl and the historical runs these traditional fingerprints do not represent either system with a high correlation. For the MSE score, we can note minimal error for the RR AMOC fingerprint compared to true AMOC, and larger MSE values for the traditional fingerprints. This is persistently the case in the historical runs as well.

\begin{figure}[H]
\centering
\begin{minipage}{1\textwidth}
    \centering
    \includegraphics[width=\textwidth]{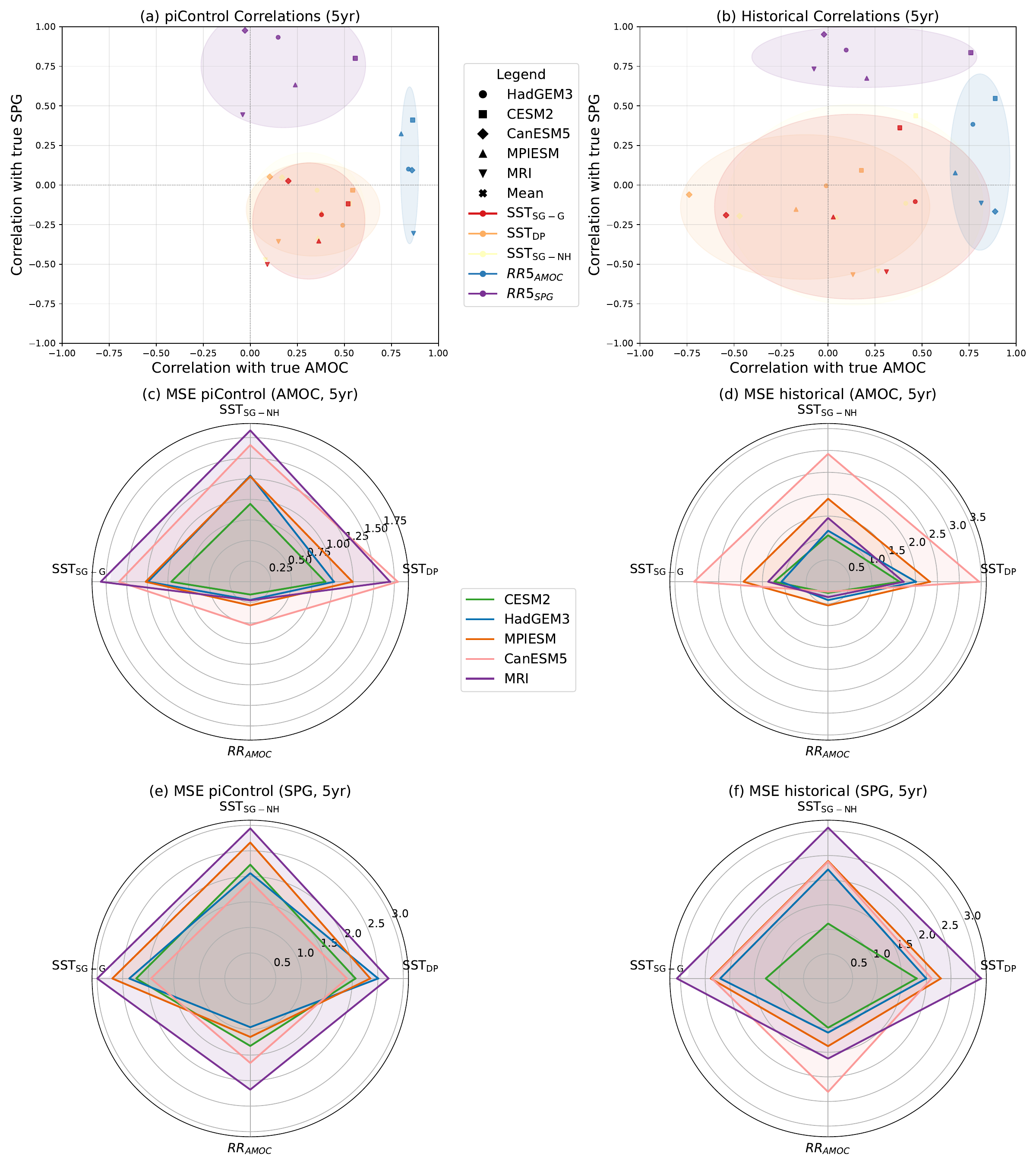}
    \caption{\textbf{Full comparison of the performance of all fingerprints in the piControl and historical runs} Panels (a) and (b) showcase Pearson correlation coefficients between each fingerprint and the true AMOC on the x-axis and the true SPG on the y-axis in the piControl and historical timeseries, respectively. Different colours represent the different fingerprints and symbols represent the different models. The ellipses are drawn to guide the eye to the clustering of the fingerprints. Blue represents the RR SPG fingerprint, purple RR AMOC fingerprint, brown represents $\mathrm{SST_{SG-G}}$ fingerprint, red $\mathrm{SST_{DP}}$ and yellow represents $\mathrm{SST_{SG-NH}}$ fingerprint. The symbols are the same as in Figure 4. Panels (c) and (d) show radar plots of MSE between each fingerprint and the true AMOC for all models. Blue represents HadGEM, orange MPI, green CESM, red CanESM and purple MRI. The top part of the panel is the $\mathrm{SST_{SG-NH}}$ fingerprint, bottom part of the panel is the RR5 AMOC fingerprint, left part of the panel is for the $\mathrm{SST_{SG-G}}$ fingerprint and the right part of the panel is for the $\mathrm{SST_{DP}}$ fingerprint. We can see that for both runs and all models the RR5 AMOC fingerprint has the lowest MSE. 
    Panels (e) and (f) show the radar plots of MSE between each fingerprint and the true SPG for all models. Colour scheme and the positioning of the fingerprints along the radar plot is the same as before. 
}
    \label{fig:fingerprint_panels}
\end{minipage}%

\end{figure}

\newpage

\section*{Discussion and Conclusion}


In this study, we utilized ridge regression (RR) to estimate the AMOC and SPG strength from observable sea surface parameters; SST and SSS, hence proposing two statistically optimal set of fingerprints, one for the AMOC and the other for the SPG. The traditionally available and widely used SST-based AMOC fingerprints are compared with our RR fingerprints for the AMOC and SPG. This comparison reveals that the traditional fingerprints do not reliably capture variability in either system, and that they provide insufficient distinction between AMOC and SPG. 

In contrast, the RR fingerprints significantly outperform traditional SST-based AMOC fingerprints in both correlation and MSE across a suite of CMIP6 models in the piControl and historical runs. This suggests that traditional fingerprints underutilize surface information that could be leveraged to better represent ocean circulation variability and trends. 

Traditional AMOC fingerprints \cite{caesar2018, jackson2020, rahmstorf2015} were primarily developed to capture long-term trends in AMOC strength. While they show moderate positive correlations with the true AMOC in the 1-30 year rolling means of the piControl runs, their performance declines notably in the historical simulations, despite the presence of some trend in historical simulations compared to none in piControl runs. This has already been shown for some of these fingerprints in previous works using CMIP6 models \cite{Menary2020,zhu2023,terhaar2025}, and some possible explanations of the poor fingerprint performance include the enhanced aerosol-cooling in CMIP6 \cite{Menary2020} or the models' inherent multicentennial variability \cite{Bonnet2021, Mehling2024}. 

Beyond trend detection, the low correlation of traditional fingerprints with the true AMOC between one to thirty year rolling means (as seen in Figure 2 and corresponding SI figures), together with their relatively high MSE, points to a limited ability to capture AMOC variability. This is a particular concern given that these same fingerprints are routinely used to detect critical slowing down as an early-warning signal of stability loss: our results suggest they might be less suitable for this purpose than previously thought, and that data-driven alternatives such as our RR-based fingerprints offer a more robust tool.

This shortcoming is compounded by the assumptions underlying several traditional fingerprints, which implicitly rely on a tight coupling between AMOC strength and SST in the SPG region. In particular, SST-based fingerprints such as the "warming hole" in the subpolar North Atlantic \cite{rahmstorf2015, caesar2018} interpret regional cooling as a sign of AMOC weakening, relying on the premise that AMOC reduction leads to reduced heat transport and hence cooling in the SPG region. However, a warming hole has been found in a slab ocean model without ocean heat transport, which may indicate that the warming hole is partly driven by atmospheric processes \cite{he2022}. Additionally, even if the warming hole is associated with AMOC weakening, recent work has shown that this connection may be only partial, and that it could manifest through different mechanisms \cite{Fan2025}.

Consequently, this assumed coupling does not consistently hold when applied to natural variability, as we demonstrate in unforced (piControl) model simulations. The fingerprints rooted in SPG-based SST anomalies fail to track AMOC variability and instead show weak or inconsistent correlations with AMOC strength, as well as a substantial lag. 

Importantly, our findings also indicate that traditional fingerprints may contain SPG signals, given that they in occasion have as much correlation with the SPG timeseries or the restoring rate of the true SPG timeseries as they do with the AMOC. In contrast, the RR fingerprint for the SPG exhibits high correlations with the true SPG and can distinguish AMOC from SPG clearly. This separation is critical, given that the two systems may undergo tipping independently, and so their warning signals should be calculated independently \cite{sgubin2017, drijfhout2015}.

A key strength of the RR-based approach lies in its model-specific optimization. In this sense, we demonstrate the maximum potential of fingerprinting (in a linear framework). Unlike spatially averaged SST fingerprints, our RR fingerprints adapt to each model’s own variability patterns, yielding higher accuracy. However, this adaptability comes with a trade-off: limited generalisability to other models and to observational data. Moreover, since the fingerprints are trained exclusively on SST and SSS, they may miss relevant dynamics captured in subsurface or atmospheric variables, and the use of a linear framework omits nonlinear mechanisms influencing the AMOC and SPG evolution.

Applied to the HadISST observational record, traditional fingerprints show a declining annual trend that is statistically significant in two of three cases, while the corresponding restoring-rate indicator points to a significant loss of stability. The RR5 fingerprint, by contrast, shows increasing trend of the restoring rate in four out of five RR5 models, this is similar in RR1 with two RR1 models showing a statistically significant increase and one RR1 model showing a statistically significant decrease. Over the longer, 30-year smoothing timescale, this picture is less consistent, with no RR30 fingerprint showing a significant restoring-rate change, compared to one of three traditional fingerprints.  

A fingerprint that aims to serve as a diagnostic for AMOC stability, or even as a precursor for potential AMOC collapse, should correlate strongly with AMOC variability while maintaining minimal correlation with SPG variability, and vice versa for SPG-specific fingerprints. Our RR-based fingerprints meet this criterion in contrast to traditional SST-based fingerprints, producing model-specific diagnostics that maximize correlation with their respective target systems, while minimizing cross-correlation. This separation is crucial for disentangling overlapping signals and for improving detection of potential instabilities in each subsystem.

Our work underscores three key conclusions: first, fingerprint selection should align with the specific objectives of the investigation (e.g. trend detection versus detection of higher-order variability); second, it is possible to construct AMOC and SPG fingerprints from SST and SSS data that are much more effective than the currently used, traditional fingerprints; and third, our regression-based fingerprints do not suggest a weakening of the AMOC or SPG when applied to historical observations, but still support a historical stability decline. Additional studies are needed to further evaluate the generalisability and the performance of these fingerprints in observational contexts.

\section*{Methods}

\subsection*{Climate models}

We select CMIP6 models with a piControl timeseries of 1000+ years with priority given to models not having the same ocean model. Eventually, the models and their respective realisations used are: CESM2 (hereinafter  abbreviated as CESM)\cite{danabasoglu2020cesm2}, HadGEM3-GC31-LL\cite{hadgem3} realisation: r1i1p1f1 (hereinafter  abbreviated as HadGEM), CanESM5 (CanESM) \cite{canesm5}, MRI-ESM2-0\cite{mri} as MRI, MPI-ESM1-2-LR\cite{mpi} as MPI. All of these models except for MRI also have been used in the North Atlantic Hosing Model Intercomparison Project (NAHosMIP) \cite{nahosmip}. 

The data used is SSS and SST fields from each model, first regridded to the HadISST observational grid and limited to the longitudinal range of 100\,°\,W to 40\,°\,E to include all relevant seas and deep water formation sites for AMOC. All data is initially yearly averaged and then smoothed with a 5-year and a 30-year centered rolling mean in some cases as indicated. Other data used includes, as abbreviated by the Climate Model Output Rewriter (CMOR) convention, the msftbarot and msftmz or msftyz variables for each model. 

\subsection*{True AMOC and SPG time series}
AMOC is defined as the maximum of the meridional streamfunction at 40\,°\,N latitude for the top 2\,km of the ocean. SPG is defined as the minimum of the barotropic streamfunction in the North Atlantic box with boundaries at 0\,° to 70\,°\,N and 80\,°\,W to 20\,°\,E. The msftbarot parameter is also regridded to HadISST observational grid before further processing. Due to the non-uniform output of CMIP6 msftbarot parameter output, some ESMs  Both model outputs are converted to Sverdrups. All figures and correlations are quoted with the barotropic streamfunction multiplied by -1 such that a large positive number represents a stronger SPG circulation. The true AMOC and true SPG we refer to are anomalies where we subtract the time mean.

\subsection*{Ridge Regression Fingerprint}

A normalisation procedure for the SSS and SST fields is carried out by subtracting the mean and dividing by the standard deviation of each field individually. This is done separately for the piControl and the historical runs. This process ensures non-dimensionalized SSS and SST values, which is necessary for the RR we perform. 
RR is defined as: 

\begin{equation}
    \hat{\beta} = \arg \min_{\beta} \left( \sum_{i=1}^{n} \left( y_i - X_i \beta \right)^2 + \alpha \sum_{j=1}^{p} \beta_j^2 \right)
\end{equation}

where $\hat{\beta}$ is the vector of estimated coefficients, $y_i$ is the target variable (AMOC or SPG timeseries), $X_i$ represents the features matrix (SSS and SST values), $\hat{\beta}_j$ represents the coefficients of the model, $\alpha$ is the penalty term and $p$ is the number of features. In this case each SSS and SST field represents a new feature. 

The RR models for AMOC and SPG were trained on the first 80\,\% of the piControl timeseries for each climate model. The penalty term was chosen in a way that optimizes the correlation between the true and reconstructed timeseries on the testing portion of the piControl timeseries for the majority of the models. We thus chose a value of $\alpha$ = 1000. The piControl trained RR was then fed SSS and SST values from the historical run and its reconstruction is compared to the model output timeseries. This process was also repeated with 15, 20 and 30-year rolling averages for both piControl and Historical runs in order to compare the RR to the other fingerprints. We note here that these are different fingerprints due to being trained on inputs and targets that are smoothed according to each window. These fingerprints are therefore, denoted with their respective windows such as RR5 and RR30. 

The Pearson correlation coefficient between the model output AMOC and SPG are computed, as well as the correlation between the RR fingerprint for AMOC and SPG (Figure 4). 

Finally, the 5-year rolling correlations and MSE between the modelled AMOC and SPG and the fingerprint reconstructions were computed, results can be found on Figure 5. 

\subsection*{Other Fingerprints}

Commonly used SST fingerprints from the literature are computed and compared to our RR fingerprint as well as to the model output.

The $\mathrm{SST_{SG-G}}$ fingerprint is calculated as the difference between the regional mean SST (averaged from November to March over the area spanning 46\,°\,-61\,°\,N and 55\,°\,-20\,°\,W) and the global SST mean over the same time steps.

The $\mathrm{SST_{DP}}$ fingerprint is defined as the annual mean SST over the North Atlantic region (45\,°\,-80\,°\,N, 70\,°\,W-30\,°\,E) minus the annual mean SST over the same longitudinal range in the South Atlantic (45\,°\,-0\,°\,S).

Finally, the $\mathrm{SST_{SG-NH}}$ fingerprint uses the same North Atlantic region as the $\mathrm{SST_{SG-G}}$ fingerprint, but subtracts the Northern Hemisphere mean SST instead of the global mean.

\subsection*{Restoring Rates and Statistical Significance}

We estimated a restoring rate, $\lambda$, for each index based on the discretization of a linear relaxation (Ornstein--Uhlenbeck-type) process,
\begin{equation}
    \frac{dX}{dt} = -\lambda X + \eta(t),
    \label{eq:ou}
\end{equation}
where $X$ denotes the deviation of the index from its local mean and $\eta(t)$ is a stochastic forcing term. Under this model, a more negative restoring-rate coefficient corresponds to a system that returns more rapidly to equilibrium following a perturbation, whereas a value approaching zero indicates progressively slower recovery and a corresponding loss of resilience.

Prior to estimating $\lambda$, each index was normalized, smoothed with each window of running mean, and detrended by subtracting a 50-year rolling mean, removing low-frequency, non-stationary variability that would otherwise bias the regression-based estimate. 

$\lambda$ was then estimated using a sliding window of length $w=70$ years, advanced one year at a time. Within each window, the data were demeaned and a residual linear trend was removed via least-squares fitting.

2 sets of 1000 Fourier surrogates were generated; one from raw data, the other from detrended data. The input for the surrogates were the fingerprint timeseries based on the HadISST data. Detrended surrogates were used for the estimation of statistical significance in restoring rate values and the surrogates generated from raw data were used for slope significance.

\subsection*{Data Availability}

All used ESM data is available at  Centre for Environmental Data Analysis (CEDA) website\cite{esgf_ceda_cmip6}. Code is available upon request.

\newpage
\bibliography{fingerprints}

\section*{Acknowledgements}

This work was funded under the framework of international cooperation program by Deutsche Forschungsgemeinschaft (DFG) and National Research Foundation of Korea (NRF): Climate Resilience under Zero-Emission Commitment Scenarios, BO 4455/2-1 (eBer-24-62950; RS-2024-00438471). N.B. and S.B. have also received funding from the Volkswagen Stiftung and the European Union's Horizon Europe research and innovation programme under grant agreement No. 101137601 (ClimTip). This is ClimTip contribution \#X. Y.S. has also been supported by the National Research Foundation of Korea (NRF) grant funded by the Korea government (MSIT) (RS-2024-00334637). J.A.B and M.B-Y. acknowledge funding from the European Union’s Horizon 2020 research and
Innovation programme under the Marie Sklodowska-Curie grant agreement No. 956170. This project is funded by the Advanced Research + Invention Agency (ARIA)

\section*{Author contributions statement}

N.B., M.B., S.B., B.E. conceived the experiment(s),  B.E. conducted the experiment(s), B.E. analysed the results with input from all authors.  All authors co-wrote and discussed the manuscript. 

\section*{Competing interests}
Authors declare no competing interests. 

\section*{Supplementary Information}
\newpage

\renewcommand{\thefigure}{SI\arabic{figure}}
\setcounter{figure}{0}

\begin{figure}[H]
\centering
\begin{minipage}{1\textwidth}
    \centering
    \includegraphics[width=\textwidth]{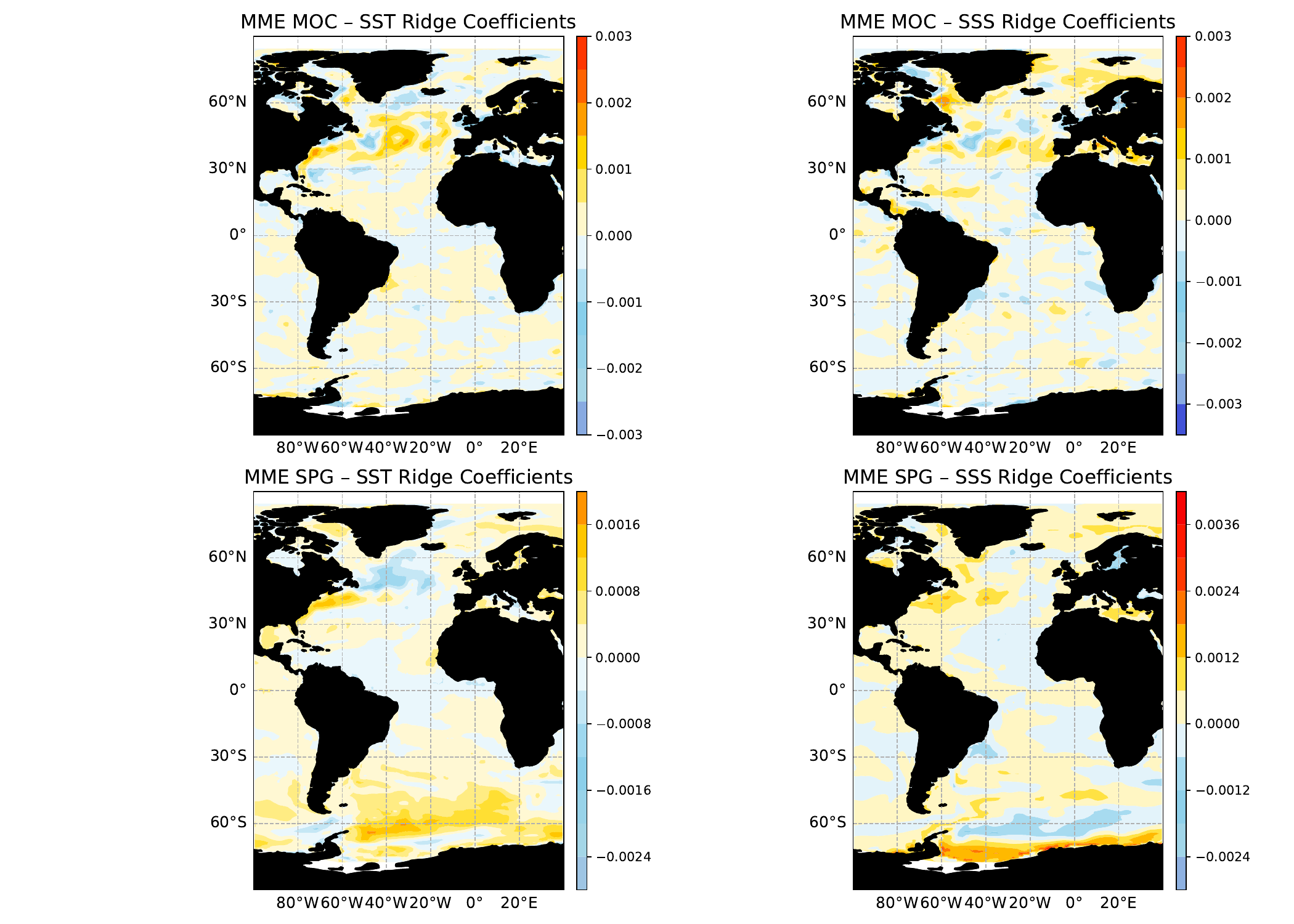}
  \captionof{figure}{\textbf{Ridge Regression Coefficient values averaged among all models} The projection of SST (left) and SSS (right) ridge coefficients as an average of all models is shown. }
    \label{fig:SI1}
\end{minipage}%

\end{figure}
\includepdf[pages=1, pagecommand={
    \thispagestyle{plain}
}]{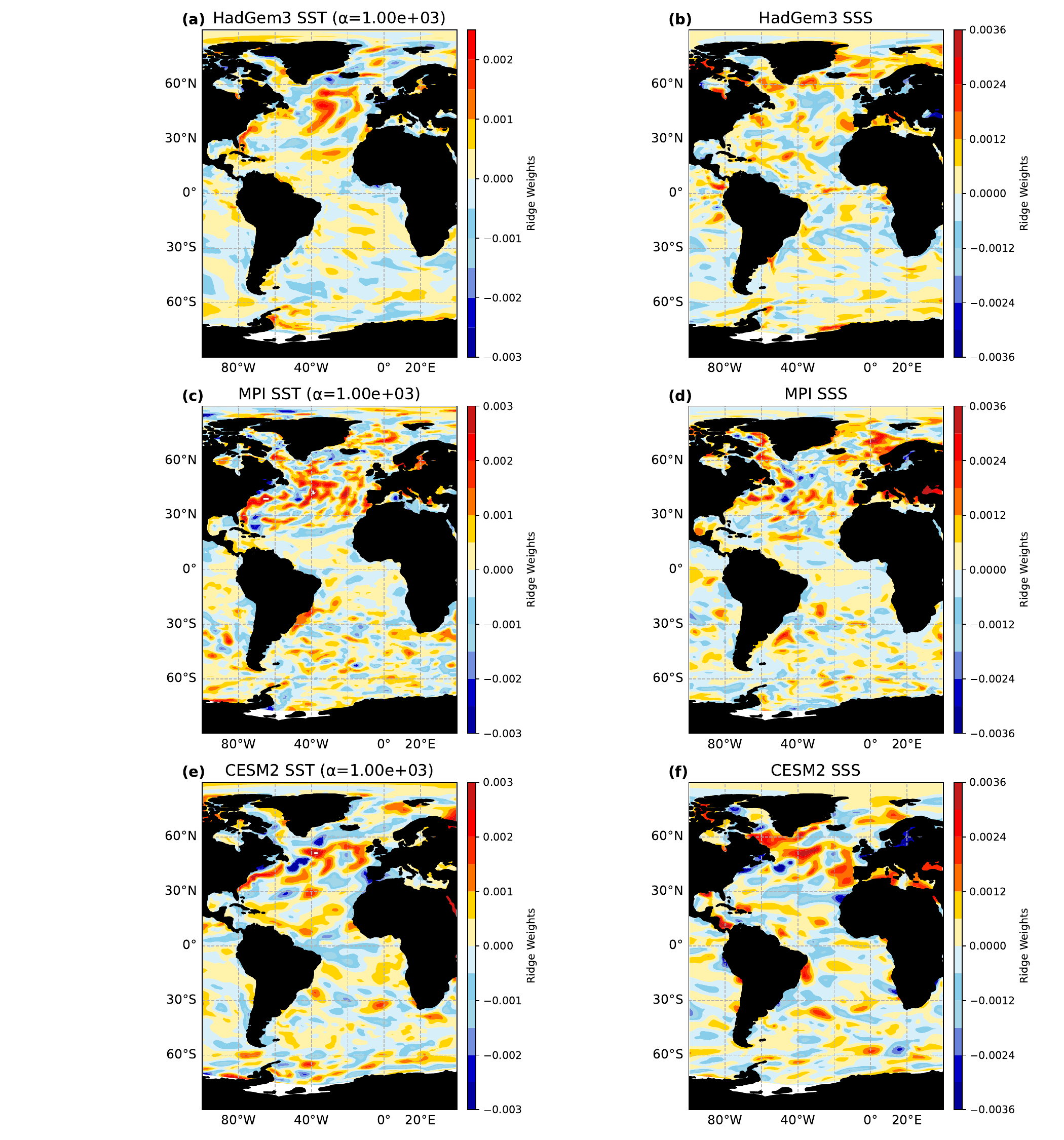}

\includepdf[pages=2, pagecommand={
  \thispagestyle{plain}
  \vspace*{\fill}
  \captionof{figure}{\textbf{RR AMOC fingerprint ridge weight maps for all models.} For each model the projection of SST (left) and SSS (right) features are shown. The weights represent the importance of a specific field for the Ridge regression reconstruction of AMOC in that model. The penalty parameter is given as 1000 for each model for ease of comparison. It can be seen that the 30-60\,°\,N - 20-40\,°\,W box is relevant for SST fields in all models for the AMOC. As for SSS, the distribution is much more varied. This figure is spread across two pages.}
  \label{fig:RR_SPG_SI1}
}]{SI/ridge_coefficients_labeled.pdf}

\includepdf[pages=1, pagecommand={
    \thispagestyle{plain}
}]{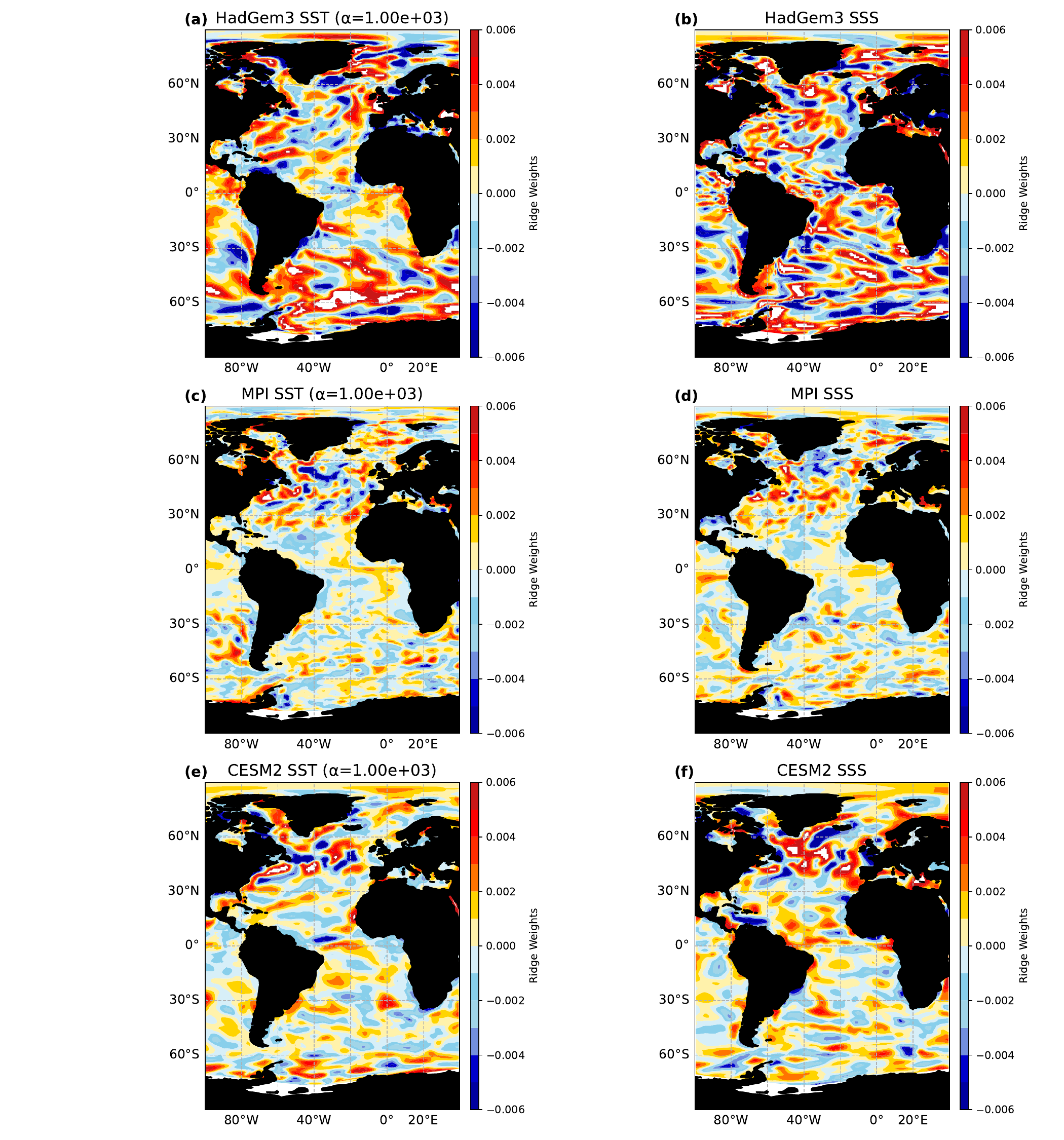}

\includepdf[pages=2, pagecommand={
  \thispagestyle{plain}
  \vspace*{\fill}
  \captionof{figure}{\textbf{RR SPG fingerprint ridge weight maps for all models.} For each model the projection of SST (left) and SSS (right) features are shown. Compared to the corresponding RR AMOC weights, the RR SPG weights are much more varied both for the SST and the SSS projections for all models. Similarly, the penalty parameter is given as 1000 for all models. This figure is spread across two pages.}
  \label{fig:RR_SPG_SI}
}]{SI/SPG_ridge_coefficients_labeled.pdf}

\begin{figure}[H]
\centering
\begin{minipage}{1\textwidth}
    \centering
    \includegraphics[width=\textwidth]{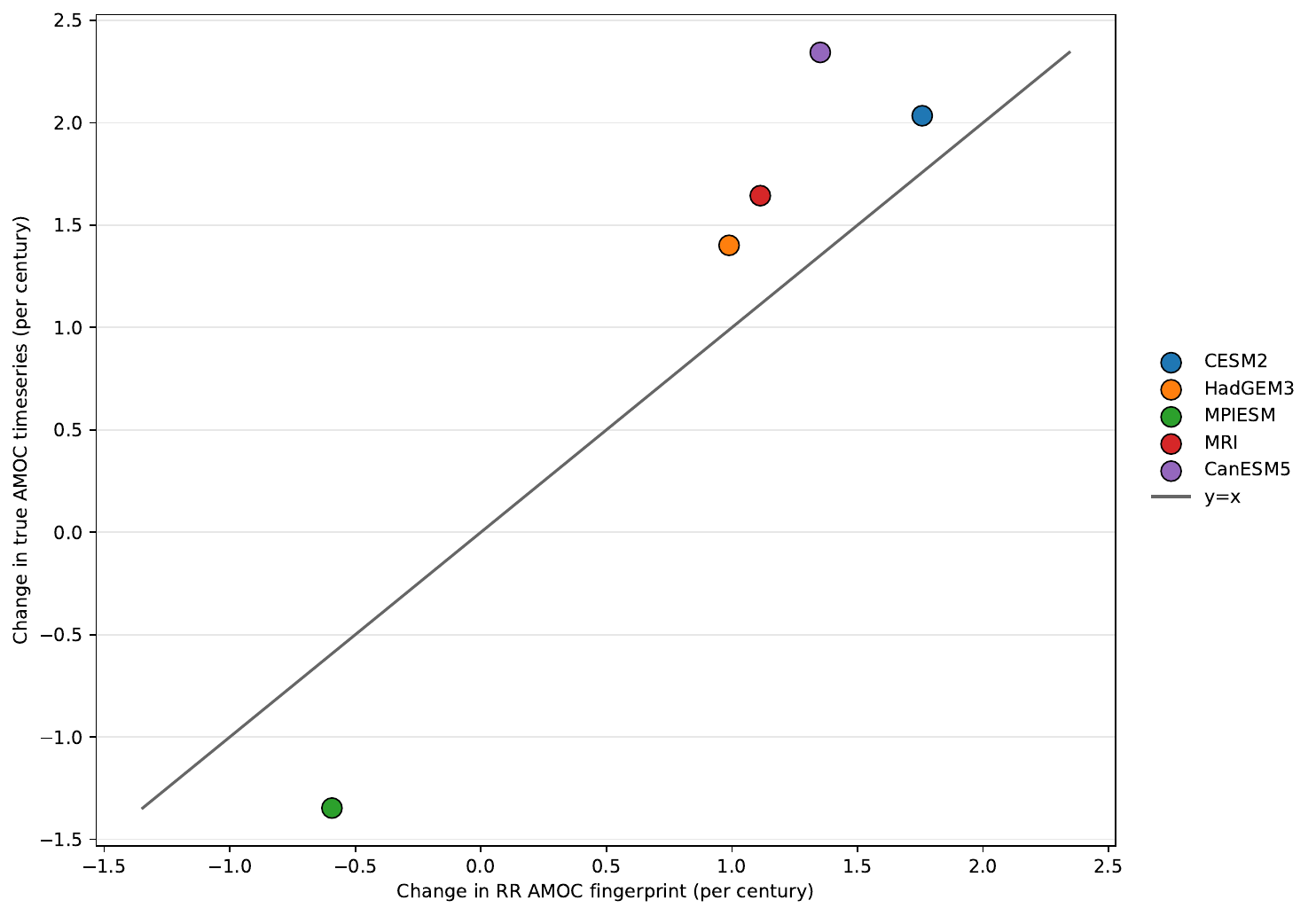}
    \caption{\textbf{Trend scatter for AMOC change per century compared to change in RR reconstruction of AMOC} The difference per century in the historical timeseries of AMOC is shown - for the true model output on the y-axis and for the RR reconstruction on the x-axis. y=x line highlights the ideal case of how the relationship between the RR fingerprint and the true AMOC should be. } 
    \label{fig:SI_scatter}
\end{minipage}

\end{figure}

\begin{figure}[H]
\centering
\begin{minipage}{1\textwidth}
    \centering
    \includegraphics[width=\textwidth]{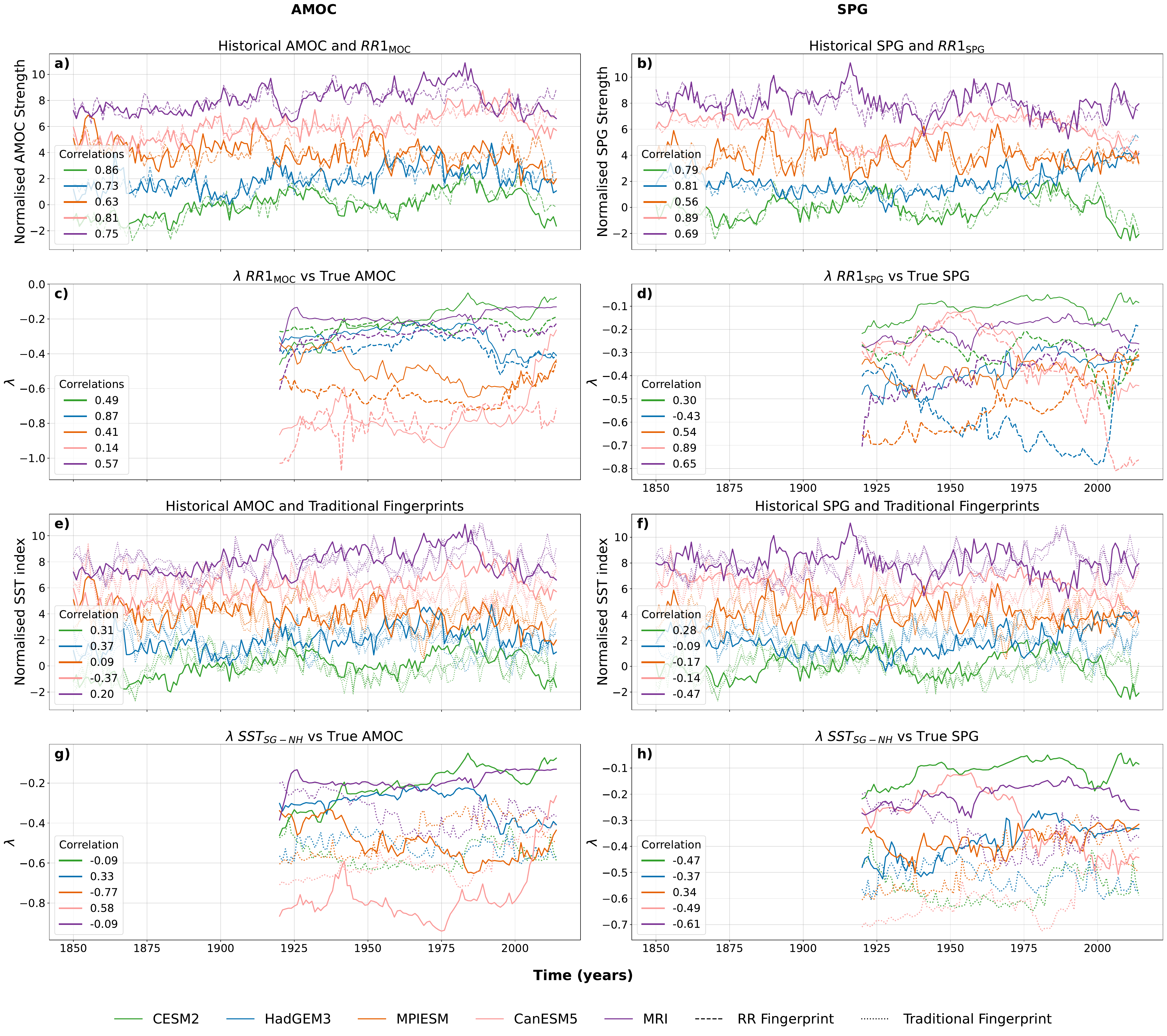}
    \caption{\textbf{Same as figure 2 but annual}}
    \label{fig:SI_fig2_1}
\end{minipage}

\end{figure}

\begin{figure}[H]
\centering
\begin{minipage}{1\textwidth}
    \centering
    \includegraphics[width=\textwidth]{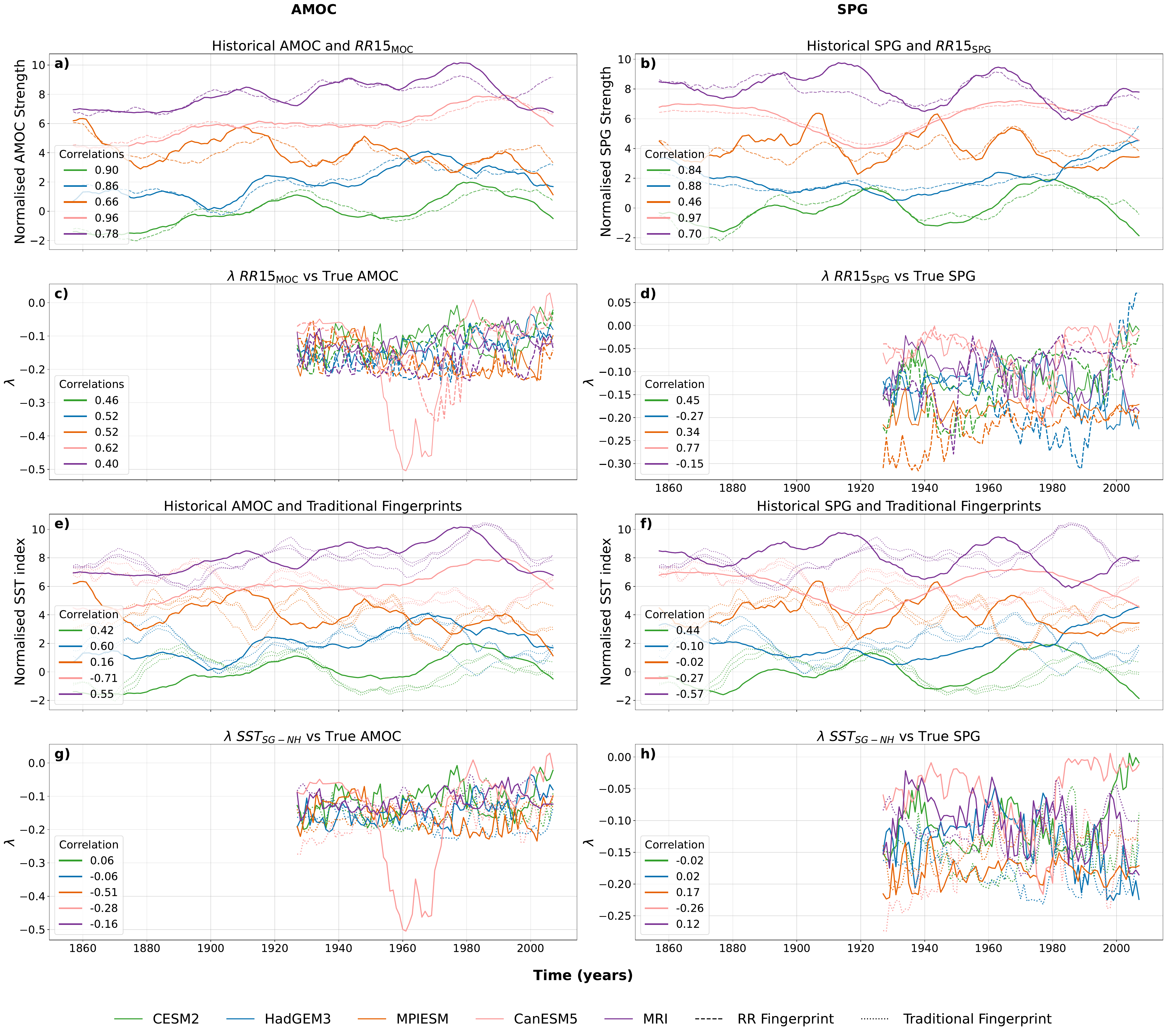}
    \caption{\textbf{Same as figure 2 but 15 year rolling mean}}
    \label{fig:SI_fig2_15}
\end{minipage}

\end{figure}

\begin{figure}[H]
\centering
\begin{minipage}{1\textwidth}
    \centering
    \includegraphics[width=\textwidth]{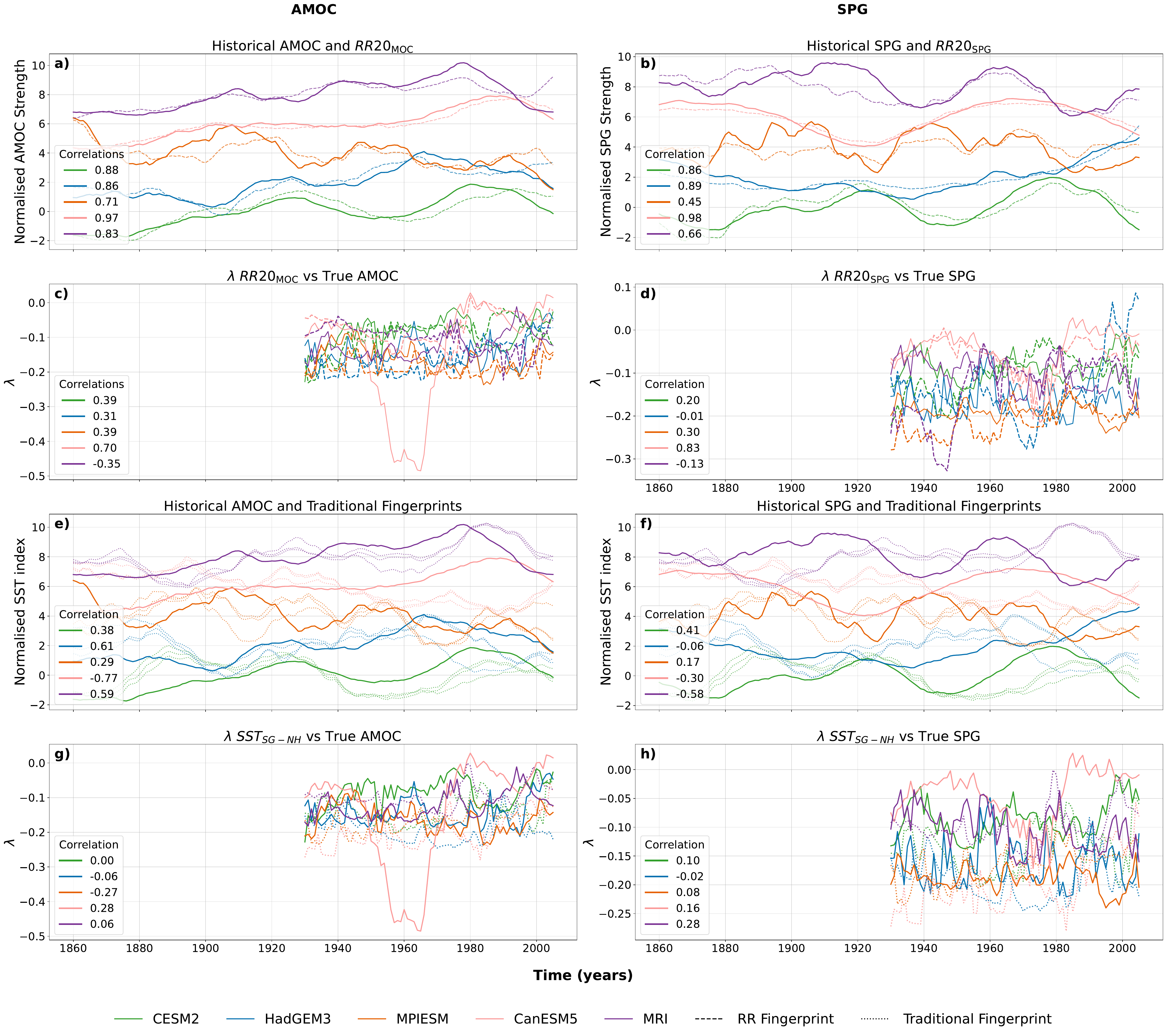}
    \caption{\textbf{Same as figure 2 but 20 year rolling mean}}
    \label{fig:SI_fig2_20}
\end{minipage}

\end{figure}

\begin{figure}[H]
\centering
\begin{minipage}{1\textwidth}
    \centering
    \includegraphics[width=\textwidth]{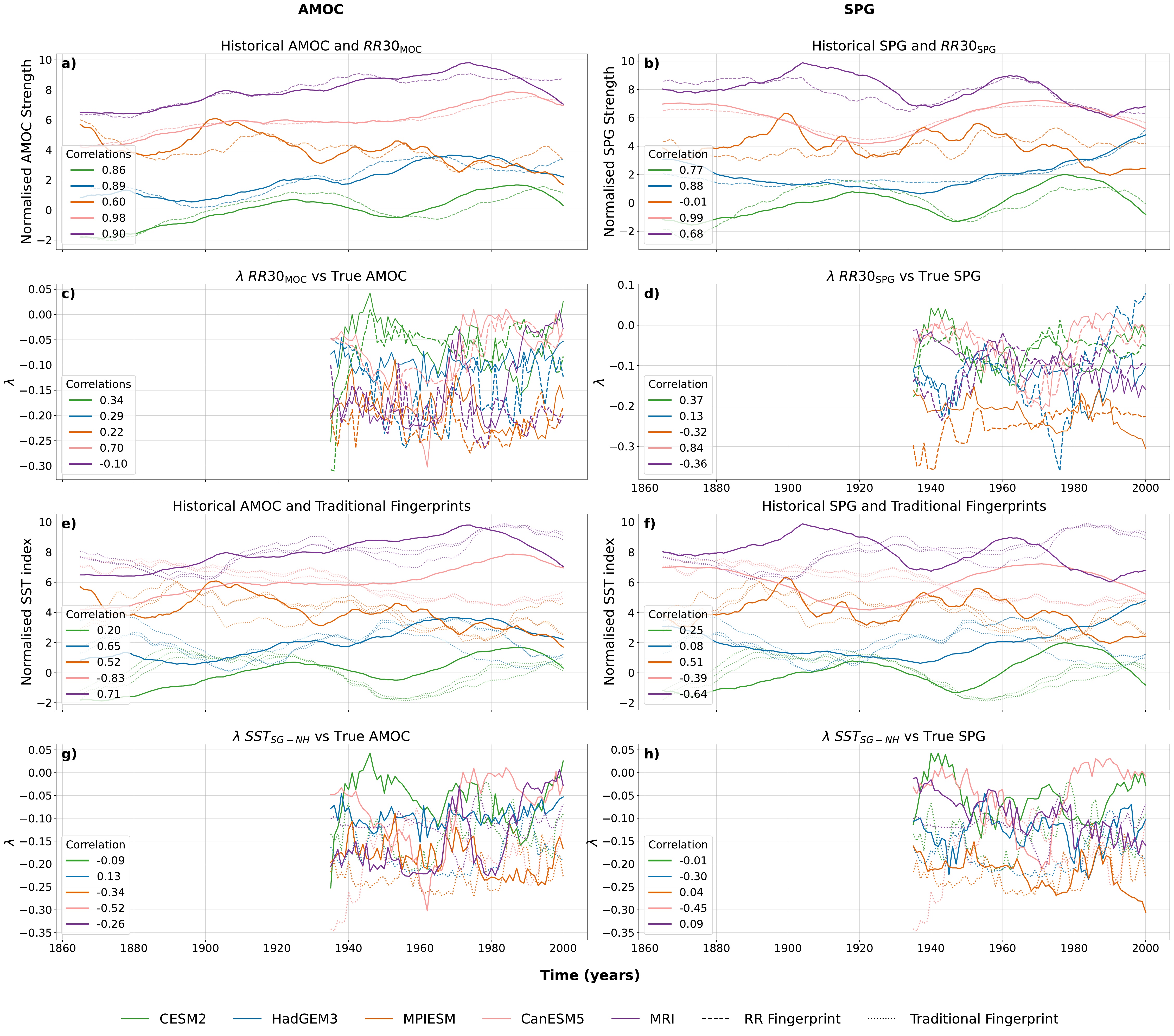}
    \caption{\textbf{Same as figure 2 but 30 year window} }
    \label{fig:SI3}
\end{minipage}

\end{figure}

\begin{figure}[H]
\centering
\begin{minipage}{1\textwidth}
    \centering
    \includegraphics[width=\textwidth]{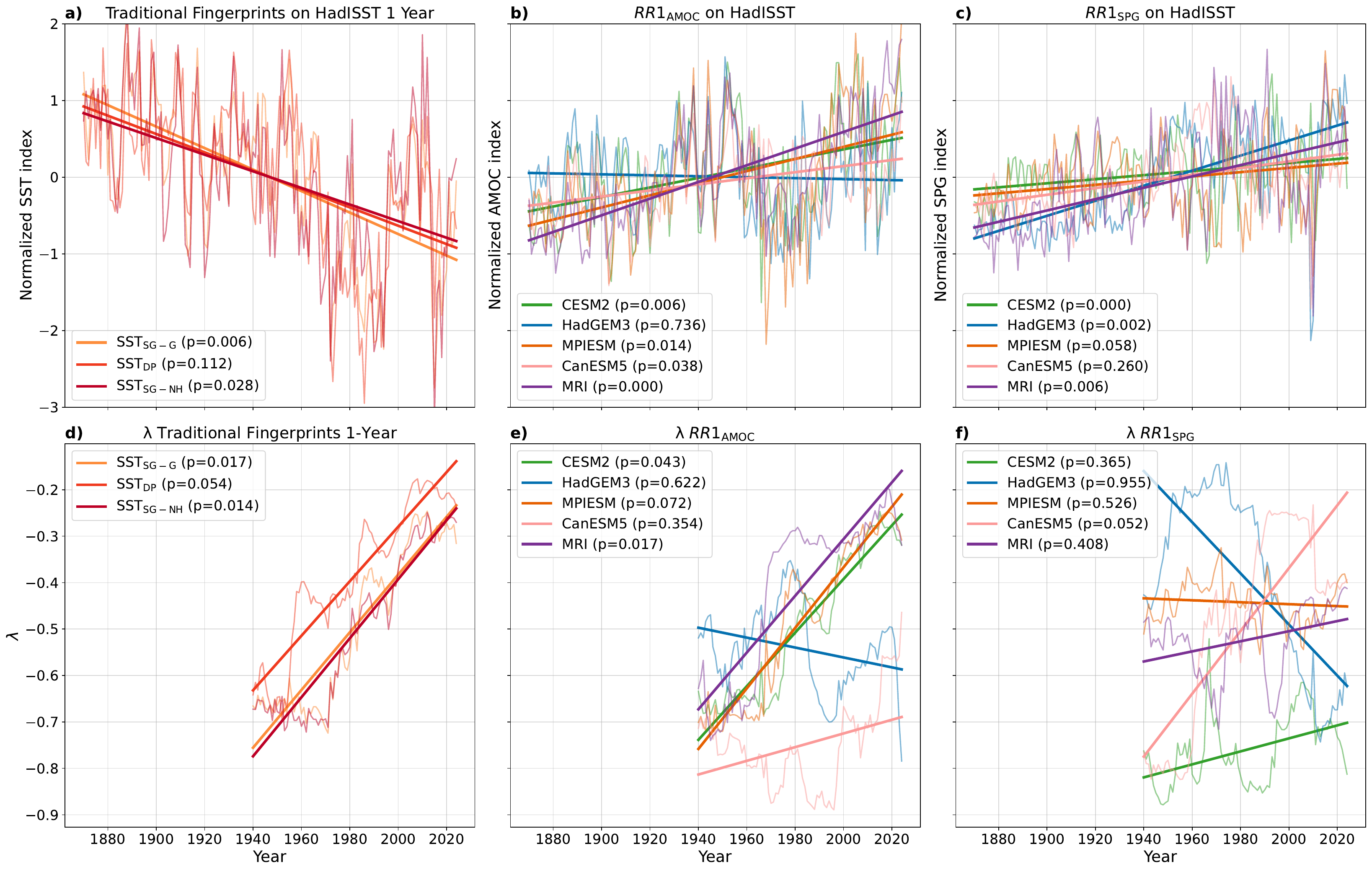}
    \caption{\textbf{Same as figure 5 but annual} }
    \label{fig:fig5_1}
\end{minipage}

\end{figure}

\begin{figure}[H]
\centering
\begin{minipage}{1\textwidth}
    \centering
    \includegraphics[width=\textwidth]{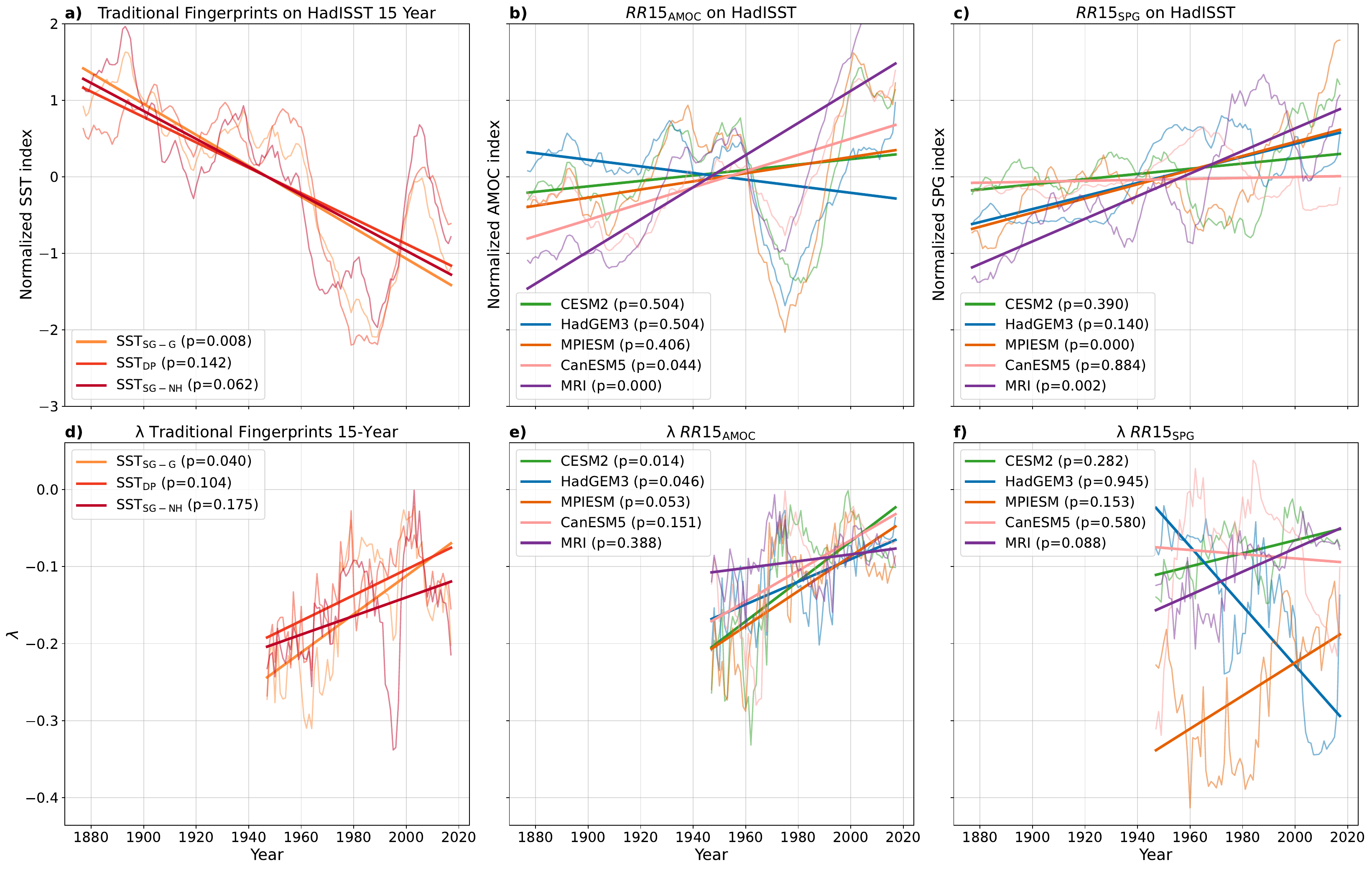}
    \caption{\textbf{Same as figure 5 but 15 year rolling mean} }
    \label{fig:fig5_15}
\end{minipage}

\end{figure}

\begin{figure}[H]
\centering
\begin{minipage}{1\textwidth}
    \centering
    \includegraphics[width=\textwidth]{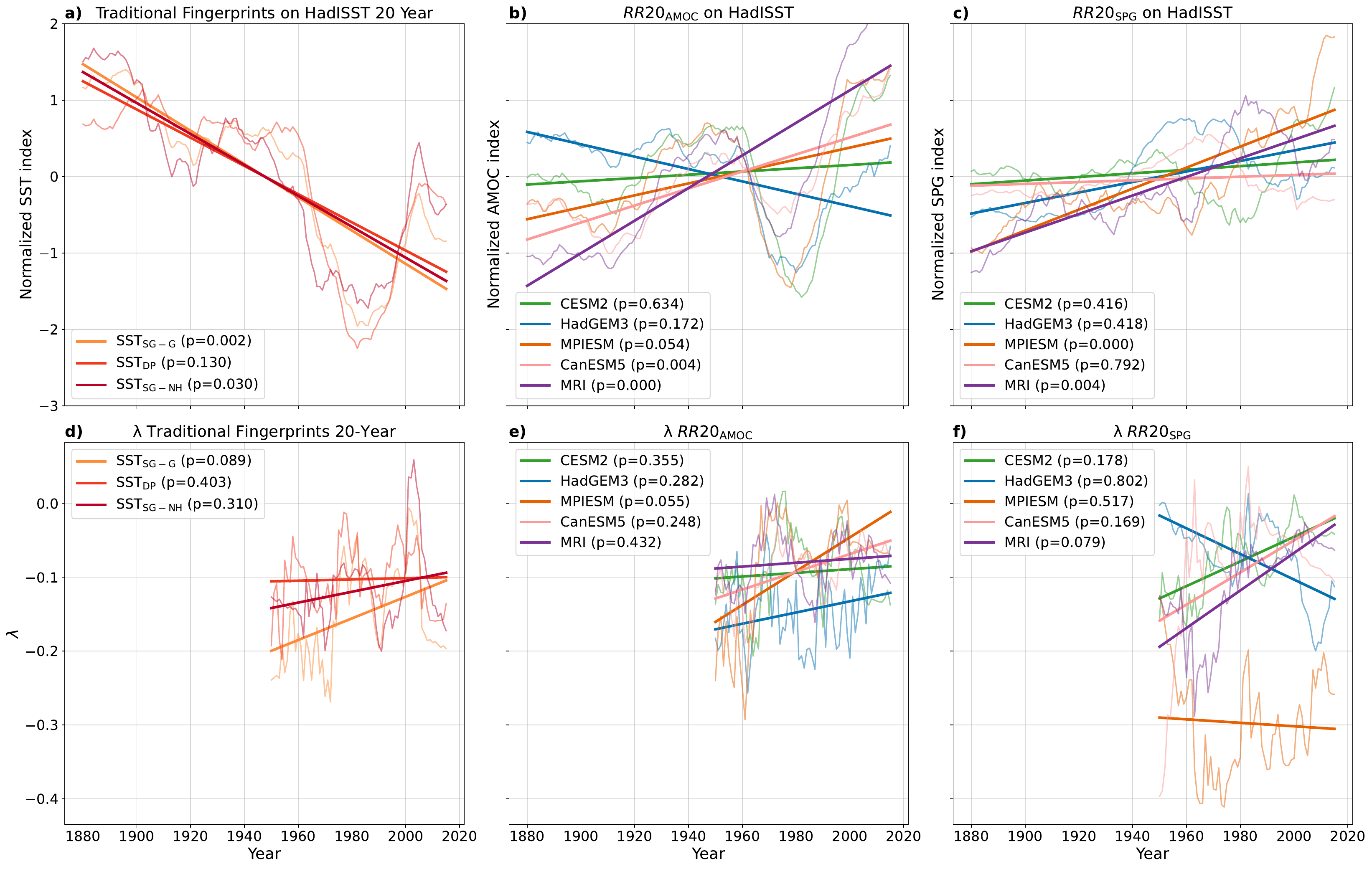}
    \caption{\textbf{Same as figure 5 but 20 year rolling mean} }
    \label{fig:fig5_20}
\end{minipage}

\end{figure}

\begin{figure}[H]
\centering
\begin{minipage}{1\textwidth}
    \centering
    \includegraphics[width=\textwidth]{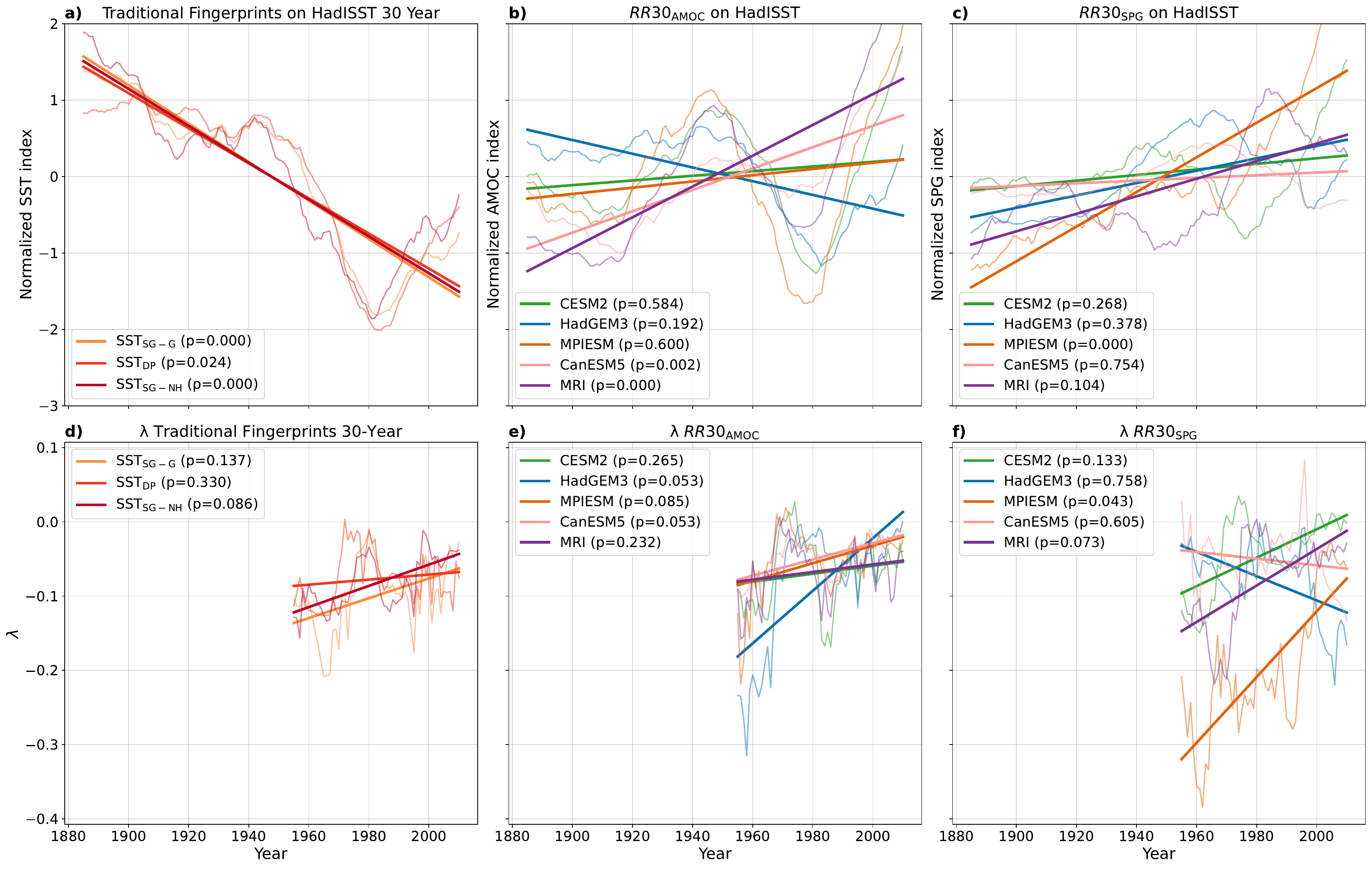}
    \caption{\textbf{Same as figure 5 but 30 year window} }
    \label{fig:SI5}
\end{minipage}

\end{figure}

\begin{table}[]
    \centering
    \caption{\textbf{Correlations of each fingerprint with respect to the true AMOC and SPG in piControl and historical timeseries.} Tables (a) and (c) show the correlations of each fingerprint with respect to the true AMOC in the respective model, as well as the mean of the models in the piControl and historical runs, respectively. Tables (b) and (d) showcase the same for the true SPG in each model. The colour-shading ranges from darker blue for values reaching -1 and darker red as the correlation grows closer to +1. The RR fingerprint for the AMOC performs best for AMOC and the RR fingerprint for the SPG performs the best for SPG. Traditional AMOC fingerprints do not have a strong positive correlation with the SPG in any model in either piControl or the historical run with the exception of the MRI historical SPG timeseries. However, they are also not highly correlated with the AMOC, with only the $\mathrm{SST_{DP}}$ fingerprint performing slightly better than the other fingerprints for AMOC in piControl timeseries. We can also see that CESM has the best performance of traditional fingerprints in the piControl. }
    \label{tab:main}

    \begin{subtable}[t]{\textwidth}
        \centering 
        \caption{Correlations with respect to AMOC in piControl; ``5 Year'' and ``30 Year'' refer to the length of the rolling mean filter applied.}
    \begin{tabular}{|l|ll|ll|ll|ll|ll|}
    \hline
                \multicolumn{1}{|c|}{\textbf{Model}} & \multicolumn{2}{c|}{\textbf{$SST_{DP}$}} & \multicolumn{2}{c|}{\textbf{$SST_{SG-NH}$}} & \multicolumn{2}{c|}{\textbf{$SST_{SG-G}$}} & \multicolumn{2}{c|}{RR SPG} & \multicolumn{2}{c|}{RR AMOC} \\ \hline

    \multicolumn{1}{|c|}{\textbf{}}      & \multicolumn{1}{c|}{\textbf{5 Year}}                                     & \multicolumn{1}{c|}{\textbf{30 Year}}               & \multicolumn{1}{c|}{\textbf{5 Year}}                                     & \multicolumn{1}{c|}{\textbf{30 Year}}               & \multicolumn{1}{c|}{\textbf{5 Year}}                                     & \multicolumn{1}{c|}{\textbf{30 Year}}               & \multicolumn{1}{c|}{\textbf{5 Year}}                                      & \multicolumn{1}{c|}{\textbf{30 Year}}                & \multicolumn{1}{c|}{\textbf{5 Year}}                                     & \multicolumn{1}{c|}{\textbf{30 Year}}               \\ \hline
    CESM2                                & \multicolumn{1}{l|}{\cellcolor[HTML]{DE735C}{\color[HTML]{F1F1F1} 0,54}} & \cellcolor[HTML]{BA2832}{\color[HTML]{F1F1F1} 0,75} & \multicolumn{1}{l|}{\cellcolor[HTML]{E17860}{\color[HTML]{F1F1F1} 0,53}} & \cellcolor[HTML]{B72230}{\color[HTML]{F1F1F1} 0,77} & \multicolumn{1}{l|}{\cellcolor[HTML]{E27B62}{\color[HTML]{F1F1F1} 0,52}} & \cellcolor[HTML]{BB2A34}{\color[HTML]{F1F1F1} 0,75} & \multicolumn{1}{l|}{\cellcolor[HTML]{DC6E57}{\color[HTML]{000000} 0,56}}  & \cellcolor[HTML]{D25849}{\color[HTML]{000000} 0,62}  & \multicolumn{1}{l|}{\cellcolor[HTML]{991027}{\color[HTML]{F1F1F1} 0,86}} & \cellcolor[HTML]{790622}{\color[HTML]{F1F1F1} 0,95} \\ \hline
    CanESM5                              & \multicolumn{1}{l|}{\cellcolor[HTML]{FAE8DE}{\color[HTML]{000000} 0,10}} & \cellcolor[HTML]{FBE4D6}{\color[HTML]{000000} 0,14} & \multicolumn{1}{l|}{\cellcolor[HTML]{FCDFCF}{\color[HTML]{000000} 0,17}} & \cellcolor[HTML]{FDD9C4}{\color[HTML]{000000} 0,21} & \multicolumn{1}{l|}{\cellcolor[HTML]{FDDBC7}{\color[HTML]{000000} 0,20}} & \cellcolor[HTML]{F9C2A7}{\color[HTML]{000000} 0,29} & \multicolumn{1}{l|}{\cellcolor[HTML]{F2F5F6}{\color[HTML]{000000} -0,03}} & \cellcolor[HTML]{F3F5F6}{\color[HTML]{000000} -0,02} & \multicolumn{1}{l|}{\cellcolor[HTML]{991027}{\color[HTML]{F1F1F1} 0,86}} & \cellcolor[HTML]{8A0B25}{\color[HTML]{F1F1F1} 0,90} \\ \hline
    HadGEM3                              & \multicolumn{1}{l|}{\cellcolor[HTML]{E6866A}{\color[HTML]{F1F1F1} 0,49}} & \cellcolor[HTML]{C94741}{\color[HTML]{F1F1F1} 0,67} & \multicolumn{1}{l|}{\cellcolor[HTML]{F6B191}{\color[HTML]{000000} 0,35}} & \cellcolor[HTML]{C94741}{\color[HTML]{F1F1F1} 0,66} & \multicolumn{1}{l|}{\cellcolor[HTML]{F5AA89}{\color[HTML]{000000} 0,38}} & \cellcolor[HTML]{C94741}{\color[HTML]{F1F1F1} 0,67} & \multicolumn{1}{l|}{\cellcolor[HTML]{FBE3D4}{\color[HTML]{000000} 0,15}}  & \cellcolor[HTML]{F4A683}{\color[HTML]{000000} 0,40}  & \multicolumn{1}{l|}{\cellcolor[HTML]{A21328}{\color[HTML]{F1F1F1} 0,84}} & \cellcolor[HTML]{790622}{\color[HTML]{F1F1F1} 0,95} \\ \hline
    MPIESM                               & \multicolumn{1}{l|}{\cellcolor[HTML]{F5AA89}{\color[HTML]{000000} 0,38}} & \cellcolor[HTML]{F5A886}{\color[HTML]{000000} 0,39} & \multicolumn{1}{l|}{\cellcolor[HTML]{F6AF8E}{\color[HTML]{000000} 0,36}} & \cellcolor[HTML]{DF765E}{\color[HTML]{F1F1F1} 0,53} & \multicolumn{1}{l|}{\cellcolor[HTML]{F6AF8E}{\color[HTML]{000000} 0,36}} & \cellcolor[HTML]{DB6B55}{\color[HTML]{F1F1F1} 0,56} & \multicolumn{1}{l|}{\cellcolor[HTML]{FBD0B9}{\color[HTML]{000000} 0,24}}  & \cellcolor[HTML]{FCE0D0}{\color[HTML]{000000} 0,16}  & \multicolumn{1}{l|}{\cellcolor[HTML]{B1182B}{\color[HTML]{F1F1F1} 0,80}} & \cellcolor[HTML]{DD7059}{\color[HTML]{F1F1F1} 0,55} \\ \hline
    MRI                                  & \multicolumn{1}{l|}{\cellcolor[HTML]{FCE2D2}{\color[HTML]{000000} 0,15}} & \cellcolor[HTML]{D55D4C}{\color[HTML]{F1F1F1} 0,61} & \multicolumn{1}{l|}{\cellcolor[HTML]{F9EBE3}{\color[HTML]{000000} 0,08}} & \cellcolor[HTML]{D7634F}{\color[HTML]{F1F1F1} 0,59} & \multicolumn{1}{l|}{\cellcolor[HTML]{FAEAE1}{\color[HTML]{000000} 0,09}} & \cellcolor[HTML]{D7634F}{\color[HTML]{F1F1F1} 0,59} & \multicolumn{1}{l|}{\cellcolor[HTML]{EFF3F5}{\color[HTML]{000000} -0,04}} & \cellcolor[HTML]{8AC0DB}{\color[HTML]{000000} -0,42} & \multicolumn{1}{l|}{\cellcolor[HTML]{960F27}{\color[HTML]{F1F1F1} 0,87}} & \cellcolor[HTML]{760521}{\color[HTML]{F1F1F1} 0,95} \\ \hline
    Mean                                 & \multicolumn{1}{l|}{\cellcolor[HTML]{F7B799}{\color[HTML]{000000} 0,33}} & \cellcolor[HTML]{E37E64}{\color[HTML]{F1F1F1} 0,51} & \multicolumn{1}{l|}{\cellcolor[HTML]{F8BFA4}{\color[HTML]{000000} 0,30}} & \cellcolor[HTML]{DD7059}{\color[HTML]{F1F1F1} 0,55} & \multicolumn{1}{l|}{\cellcolor[HTML]{F8BDA1}{\color[HTML]{000000} 0,31}} & \cellcolor[HTML]{DA6853}{\color[HTML]{F1F1F1} 0,57} & \multicolumn{1}{l|}{\cellcolor[HTML]{FCDECD}{\color[HTML]{000000} 0,18}}  & \cellcolor[HTML]{FCE2D2}{\color[HTML]{000000} 0,15}  & \multicolumn{1}{l|}{\cellcolor[HTML]{9F1228}{\color[HTML]{F1F1F1} 0,85}} & \cellcolor[HTML]{991027}{\color[HTML]{F1F1F1} 0,86} \\ \hline
    \end{tabular}
    \end{subtable}

    \vspace{1em}

    \begin{subtable}[t]{\textwidth}
        \centering
        \caption{Correlations with respect to SPG in piControl}
    \begin{tabular}{|l|ll|ll|ll|ll|ll|}
    \hline
            \multicolumn{1}{|c|}{\textbf{Model}} & \multicolumn{2}{c|}{\textbf{$SST_{DP}$}} & \multicolumn{2}{c|}{\textbf{$SST_{SG-NH}$}} & \multicolumn{2}{c|}{\textbf{$SST_{SG-G}$}} & \multicolumn{2}{c|}{RR SPG} & \multicolumn{2}{c|}{RR AMOC} \\ \hline
     
    \multicolumn{1}{|c|}{\textbf{}}      & \multicolumn{1}{c|}{\textbf{5 Year}}                                      & \multicolumn{1}{c|}{\textbf{30 Year}}                & \multicolumn{1}{c|}{\textbf{5 Year}}                                      & \multicolumn{1}{c|}{\textbf{30 Year}}                & \multicolumn{1}{c|}{\textbf{5 Year}}                                      & \multicolumn{1}{c|}{\textbf{30 Year}}                & \multicolumn{1}{c|}{\textbf{5 Year}}                                     & \multicolumn{1}{c|}{\textbf{30 Year}}               & \multicolumn{1}{c|}{\textbf{5 Year}}                                      & \multicolumn{1}{c|}{\textbf{30 Year}}                \\ \hline
    CESM2                                & \multicolumn{1}{l|}{\cellcolor[HTML]{F0F4F6}{\color[HTML]{000000} -0,03}} & \cellcolor[HTML]{F9EBE3}{\color[HTML]{000000} 0,08}  & \multicolumn{1}{l|}{\cellcolor[HTML]{E1EDF3}{\color[HTML]{000000} -0,12}} & \cellcolor[HTML]{F9EBE3}{\color[HTML]{000000} 0,08}  & \multicolumn{1}{l|}{\cellcolor[HTML]{E0ECF3}{\color[HTML]{000000} -0,12}} & \cellcolor[HTML]{F9EBE3}{\color[HTML]{000000} 0,08}  & \multicolumn{1}{l|}{\cellcolor[HTML]{B1182B}{\color[HTML]{F1F1F1} 0,80}} & \cellcolor[HTML]{BF3338}{\color[HTML]{F1F1F1} 0,72} & \multicolumn{1}{l|}{\cellcolor[HTML]{F2A17F}{\color[HTML]{330001} 0,41}}  & \cellcolor[HTML]{EB9172}{\color[HTML]{330001} 0,46}  \\ \hline
    CanESM5                              & \multicolumn{1}{l|}{\cellcolor[HTML]{F9F0EB}{\color[HTML]{000000} 0,05}}  & \cellcolor[HTML]{FBE3D4}{\color[HTML]{000000} 0,15}  & \multicolumn{1}{l|}{\cellcolor[HTML]{F9F0EB}{\color[HTML]{000000} 0,05}}  & \cellcolor[HTML]{FCE2D2}{\color[HTML]{000000} 0,15}  & \multicolumn{1}{l|}{\cellcolor[HTML]{F8F3F0}{\color[HTML]{000000} 0,03}}  & \cellcolor[HTML]{FAE7DC}{\color[HTML]{000000} 0,11}  & \multicolumn{1}{l|}{\cellcolor[HTML]{6D0220}{\color[HTML]{F1F1F1} 0,98}} & \cellcolor[HTML]{67001F}{\color[HTML]{F1F1F1} 0,99} & \multicolumn{1}{l|}{\cellcolor[HTML]{FAE9DF}{\color[HTML]{330001} 0,09}}  & \cellcolor[HTML]{FCDFCF}{\color[HTML]{330001} 0,17}  \\ \hline
    HadGEM3                              & \multicolumn{1}{l|}{\cellcolor[HTML]{C0DCEB}{\color[HTML]{000000} -0,25}} & \cellcolor[HTML]{A9D1E5}{\color[HTML]{000000} -0,32} & \multicolumn{1}{l|}{\cellcolor[HTML]{F0F4F6}{\color[HTML]{000000} -0,03}} & \cellcolor[HTML]{F8F4F2}{\color[HTML]{000000} 0,02}  & \multicolumn{1}{l|}{\cellcolor[HTML]{D4E6F1}{\color[HTML]{000000} -0,19}} & \cellcolor[HTML]{C2DDEC}{\color[HTML]{000000} -0,25} & \multicolumn{1}{l|}{\cellcolor[HTML]{7F0823}{\color[HTML]{F1F1F1} 0,93}} & \cellcolor[HTML]{810823}{\color[HTML]{F1F1F1} 0,92} & \multicolumn{1}{l|}{\cellcolor[HTML]{FAE9DF}{\color[HTML]{330001} 0,10}}  & \cellcolor[HTML]{E37E64}{\color[HTML]{330001} 0,51}  \\ \hline
    MPIESM                               & \multicolumn{1}{l|}{\cellcolor[HTML]{D5E7F1}{\color[HTML]{000000} -0,18}} & \cellcolor[HTML]{EFF3F5}{\color[HTML]{000000} -0,04} & \multicolumn{1}{l|}{\cellcolor[HTML]{A7D0E4}{\color[HTML]{000000} -0,33}} & \cellcolor[HTML]{D1E5F0}{\color[HTML]{000000} -0,20} & \multicolumn{1}{l|}{\cellcolor[HTML]{A0CCE2}{\color[HTML]{000000} -0,35}} & \cellcolor[HTML]{DAE9F2}{\color[HTML]{000000} -0,15} & \multicolumn{1}{l|}{\cellcolor[HTML]{D05548}{\color[HTML]{F1F1F1} 0,63}} & \cellcolor[HTML]{FCE2D2}{\color[HTML]{000000} 0,15} & \multicolumn{1}{l|}{\cellcolor[HTML]{F7B99C}{\color[HTML]{330001} 0,32}}  & \cellcolor[HTML]{F5F6F7}{\color[HTML]{330001} -0,01} \\ \hline
    MRI                                  & \multicolumn{1}{l|}{\cellcolor[HTML]{A0CCE2}{\color[HTML]{000000} -0,36}} & \cellcolor[HTML]{E7F0F4}{\color[HTML]{000000} -0,08} & \multicolumn{1}{l|}{\cellcolor[HTML]{75B2D4}{\color[HTML]{000000} -0,47}} & \cellcolor[HTML]{D2E6F0}{\color[HTML]{000000} -0,19} & \multicolumn{1}{l|}{\cellcolor[HTML]{68ABD0}{\color[HTML]{000000} -0,50}} & \cellcolor[HTML]{C5DFEC}{\color[HTML]{000000} -0,24} & \multicolumn{1}{l|}{\cellcolor[HTML]{EE9677}{\color[HTML]{000000} 0,44}} & \cellcolor[HTML]{F5AA89}{\color[HTML]{000000} 0,38} & \multicolumn{1}{l|}{\cellcolor[HTML]{AED3E6}{\color[HTML]{330001} -0,30}} & \cellcolor[HTML]{3A87BD}{\color[HTML]{330001} -0,66} \\ \hline
    Mean                                 & \multicolumn{1}{l|}{\cellcolor[HTML]{DAE9F2}{\color[HTML]{000000} -0,15}} & \cellcolor[HTML]{EFF3F5}{\color[HTML]{000000} -0,04} & \multicolumn{1}{l|}{\cellcolor[HTML]{D4E6F1}{\color[HTML]{000000} -0,18}} & \cellcolor[HTML]{F2F5F6}{\color[HTML]{000000} -0,03} & \multicolumn{1}{l|}{\cellcolor[HTML]{C7E0ED}{\color[HTML]{000000} -0,23}} & \cellcolor[HTML]{E6EFF4}{\color[HTML]{000000} -0,09} & \multicolumn{1}{l|}{\cellcolor[HTML]{BA2832}{\color[HTML]{F1F1F1} 0,76}} & \cellcolor[HTML]{CF5246}{\color[HTML]{F1F1F1} 0,63} & \multicolumn{1}{l|}{\cellcolor[HTML]{FBE6DA}{\color[HTML]{330001} 0,12}}  & \cellcolor[HTML]{FAEAE1}{\color[HTML]{330001} 0,09}  \\ \hline
    \end{tabular}
    \end{subtable}
    \vspace{1em}

    \begin{subtable}[t]{\textwidth}
        \centering
        \caption{Correlations with respect to AMOC in historical}
        \begin{tabular}{|l|ll|ll|ll|ll|ll|}
            \hline
            \multicolumn{1}{|c|}{\textbf{Model}} & \multicolumn{2}{c|}{\textbf{$SST_{DP}$}} & \multicolumn{2}{c|}{\textbf{$SST_{SG-NH}$}} & \multicolumn{2}{c|}{\textbf{$SST_{SG-G}$}} & \multicolumn{2}{c|}{RR SPG} & \multicolumn{2}{c|}{RR AMOC} \\ \hline
          
        \multicolumn{1}{|c|}{\textbf{}}      & \multicolumn{1}{c|}{\textbf{5 Year}}                                      & \multicolumn{1}{c|}{\textbf{30 Year}}                & \multicolumn{1}{c|}{\textbf{5 Year}}                                      & \multicolumn{1}{c|}{\textbf{30 Year}}                & \multicolumn{1}{c|}{\textbf{5 Year}}                                      & \multicolumn{1}{c|}{\textbf{30 Year}}                & \multicolumn{1}{c|}{\textbf{5 Year}}                                      & \multicolumn{1}{c|}{\textbf{30 Year}}                & \multicolumn{1}{c|}{\textbf{5 Year}}                                     & \multicolumn{1}{c|}{\textbf{30 Year}}               \\ \hline
        CESM2                                & \multicolumn{1}{l|}{\cellcolor[HTML]{FCDECD}{\color[HTML]{000000} 0,18}}  & \cellcolor[HTML]{E4EEF4}{\color[HTML]{000000} -0,09} & \multicolumn{1}{l|}{\cellcolor[HTML]{EA8E70}{\color[HTML]{F1F1F1} 0,47}}  & \cellcolor[HTML]{F5A886}{\color[HTML]{000000} 0,39}  & \multicolumn{1}{l|}{\cellcolor[HTML]{F5AA89}{\color[HTML]{000000} 0,38}}  & \cellcolor[HTML]{FDDBC7}{\color[HTML]{000000} 0,20}  & \multicolumn{1}{l|}{\cellcolor[HTML]{B82531}{\color[HTML]{F1F1F1} 0,76}}  & \cellcolor[HTML]{B61F2E}{\color[HTML]{F1F1F1} 0,78}  & \multicolumn{1}{l|}{\cellcolor[HTML]{900D26}{\color[HTML]{F1F1F1} 0,89}} & \cellcolor[HTML]{991027}{\color[HTML]{F1F1F1} 0,86} \\ \hline
        CanESM5                              & \multicolumn{1}{l|}{\cellcolor[HTML]{2B73B3}{\color[HTML]{F1F1F1} -0,74}} & \cellcolor[HTML]{0F437B}{\color[HTML]{F1F1F1} -0,92} & \multicolumn{1}{l|}{\cellcolor[HTML]{75B2D4}{\color[HTML]{F1F1F1} -0,47}} & \cellcolor[HTML]{2267AC}{\color[HTML]{F1F1F1} -0,79} & \multicolumn{1}{l|}{\cellcolor[HTML]{59A1CA}{\color[HTML]{F1F1F1} -0,54}} & \cellcolor[HTML]{1D5FA2}{\color[HTML]{F1F1F1} -0,83} & \multicolumn{1}{l|}{\cellcolor[HTML]{F3F5F6}{\color[HTML]{000000} -0,02}} & \cellcolor[HTML]{FAEAE1}{\color[HTML]{000000} 0,09}  & \multicolumn{1}{l|}{\cellcolor[HTML]{900D26}{\color[HTML]{F1F1F1} 0,89}} & \cellcolor[HTML]{6D0220}{\color[HTML]{F1F1F1} 0,98} \\ \hline
        HadGEM3                              & \multicolumn{1}{l|}{\cellcolor[HTML]{F5F6F7}{\color[HTML]{000000} -0,01}} & \cellcolor[HTML]{A2CDE3}{\color[HTML]{000000} -0,35} & \multicolumn{1}{l|}{\cellcolor[HTML]{F2A17F}{\color[HTML]{000000} 0,41}}  & \cellcolor[HTML]{D05548}{\color[HTML]{F1F1F1} 0,63}  & \multicolumn{1}{l|}{\cellcolor[HTML]{EA8E70}{\color[HTML]{F1F1F1} 0,46}}  & \cellcolor[HTML]{CC4C44}{\color[HTML]{F1F1F1} 0,65}  & \multicolumn{1}{l|}{\cellcolor[HTML]{FAE9DF}{\color[HTML]{000000} 0,10}}  & \cellcolor[HTML]{E6866A}{\color[HTML]{F1F1F1} 0,49}  & \multicolumn{1}{l|}{\cellcolor[HTML]{B72230}{\color[HTML]{F1F1F1} 0,77}} & \cellcolor[HTML]{8D0C25}{\color[HTML]{F1F1F1} 0,89} \\ \hline
        MPIESM                               & \multicolumn{1}{l|}{\cellcolor[HTML]{D7E8F1}{\color[HTML]{000000} -0,17}} & \cellcolor[HTML]{F8F2EF}{\color[HTML]{000000} 0,04}  & \multicolumn{1}{l|}{\cellcolor[HTML]{F8F1ED}{\color[HTML]{000000} 0,04}}  & \cellcolor[HTML]{DD7059}{\color[HTML]{F1F1F1} 0,55}  & \multicolumn{1}{l|}{\cellcolor[HTML]{F8F3F0}{\color[HTML]{000000} 0,03}}  & \cellcolor[HTML]{E17860}{\color[HTML]{F1F1F1} 0,52}  & \multicolumn{1}{l|}{\cellcolor[HTML]{FDD9C4}{\color[HTML]{000000} 0,21}}  & \cellcolor[HTML]{B3D6E8}{\color[HTML]{000000} -0,30} & \multicolumn{1}{l|}{\cellcolor[HTML]{C84440}{\color[HTML]{F1F1F1} 0,68}} & \cellcolor[HTML]{D6604D}{\color[HTML]{F1F1F1} 0,60} \\ \hline
        MRI                                  & \multicolumn{1}{l|}{\cellcolor[HTML]{FBE5D8}{\color[HTML]{000000} 0,13}}  & \cellcolor[HTML]{F5A886}{\color[HTML]{000000} 0,38}  & \multicolumn{1}{l|}{\cellcolor[HTML]{FAC8AF}{\color[HTML]{000000} 0,27}}  & \cellcolor[HTML]{CB4942}{\color[HTML]{F1F1F1} 0,66}  & \multicolumn{1}{l|}{\cellcolor[HTML]{F8BDA1}{\color[HTML]{000000} 0,31}}  & \cellcolor[HTML]{C13639}{\color[HTML]{F1F1F1} 0,71}  & \multicolumn{1}{l|}{\cellcolor[HTML]{E9F0F4}{\color[HTML]{000000} -0,07}} & \cellcolor[HTML]{A9D1E5}{\color[HTML]{000000} -0,32} & \multicolumn{1}{l|}{\cellcolor[HTML]{AB162A}{\color[HTML]{F1F1F1} 0,81}} & \cellcolor[HTML]{8A0B25}{\color[HTML]{F1F1F1} 0,90} \\ \hline
        Mean                                 & \multicolumn{1}{l|}{\cellcolor[HTML]{E0ECF3}{\color[HTML]{000000} -0,12}} & \cellcolor[HTML]{D2E6F0}{\color[HTML]{000000} -0,19} & \multicolumn{1}{l|}{\cellcolor[HTML]{FBE3D4}{\color[HTML]{000000} 0,14}}  & \cellcolor[HTML]{F9C4A9}{\color[HTML]{000000} 0,29}  & \multicolumn{1}{l|}{\cellcolor[HTML]{FBE5D8}{\color[HTML]{000000} 0,13}}  & \cellcolor[HTML]{FBCCB4}{\color[HTML]{000000} 0,25}  & \multicolumn{1}{l|}{\cellcolor[HTML]{FDDCC9}{\color[HTML]{000000} 0,19}}  & \cellcolor[HTML]{FBE3D4}{\color[HTML]{000000} 0,15}  & \multicolumn{1}{l|}{\cellcolor[HTML]{AE172A}{\color[HTML]{F1F1F1} 0,81}} & \cellcolor[HTML]{9F1228}{\color[HTML]{F1F1F1} 0,85} \\ \hline
        \end{tabular}
    \end{subtable}
    \vspace{1em}

    \begin{subtable}[t]{\textwidth}
        \centering
        \caption{Correlations with respect to SPG in historical}
        \begin{tabular}{|l|ll|ll|ll|ll|ll|}
            \hline
            \multicolumn{1}{|c|}{\textbf{Model}} & \multicolumn{2}{c|}{\textbf{$SST_{DP}$}}                                                 & \multicolumn{2}{c|}{\textbf{$SST_{SG-NH}$}}                                                    & \multicolumn{2}{c|}{\textbf{$SST_{SG-G}$}}                                                 & \multicolumn{2}{c|}{RR   SPG}                                                 & \multicolumn{2}{c|}{RR   AMOC}                                                 \\ \hline
                
    \multicolumn{1}{|c|}{\textbf{}}      & \multicolumn{1}{c|}{\textbf{5 Year}}                                      & \multicolumn{1}{c|}{\textbf{30 Year}}                & \multicolumn{1}{c|}{\textbf{5 Year}}                                      & \multicolumn{1}{c|}{\textbf{30 Year}}                & \multicolumn{1}{c|}{\textbf{5 Year}}                                      & \multicolumn{1}{c|}{\textbf{30 Year}}                & \multicolumn{1}{c|}{\textbf{5 Year}}                                     & \multicolumn{1}{c|}{\textbf{30 Year}}                & \multicolumn{1}{c|}{\textbf{5 Year}}                                      & \multicolumn{1}{c|}{\textbf{30 Year}}                \\ \hline
    CESM2                                & \multicolumn{1}{l|}{\cellcolor[HTML]{FAEAE1}{\color[HTML]{000000} 0,09}}  & \cellcolor[HTML]{ECF2F5}{\color[HTML]{000000} -0,06} & \multicolumn{1}{l|}{\cellcolor[HTML]{EF9979}{\color[HTML]{000000} 0,44}}  & \cellcolor[HTML]{F09C7B}{\color[HTML]{000000} 0,43}  & \multicolumn{1}{l|}{\cellcolor[HTML]{F6AF8E}{\color[HTML]{000000} 0,36}}  & \cellcolor[HTML]{FBCCB4}{\color[HTML]{000000} 0,25}  & \multicolumn{1}{l|}{\cellcolor[HTML]{A21328}{\color[HTML]{F1F1F1} 0,84}} & \cellcolor[HTML]{B72230}{\color[HTML]{F1F1F1} 0,77}  & \multicolumn{1}{l|}{\cellcolor[HTML]{DD7059}{\color[HTML]{000000} 0,55}}  & \cellcolor[HTML]{D55D4C}{\color[HTML]{000000} 0,60}  \\ \hline
    CanESM5                              & \multicolumn{1}{l|}{\cellcolor[HTML]{ECF2F5}{\color[HTML]{000000} -0,06}} & \cellcolor[HTML]{DDEBF2}{\color[HTML]{000000} -0,14} & \multicolumn{1}{l|}{\cellcolor[HTML]{D1E5F0}{\color[HTML]{000000} -0,20}} & \cellcolor[HTML]{98C8E0}{\color[HTML]{000000} -0,38} & \multicolumn{1}{l|}{\cellcolor[HTML]{D2E6F0}{\color[HTML]{000000} -0,19}} & \cellcolor[HTML]{96C7DF}{\color[HTML]{000000} -0,39} & \multicolumn{1}{l|}{\cellcolor[HTML]{790622}{\color[HTML]{F1F1F1} 0,95}} & \cellcolor[HTML]{6A011F}{\color[HTML]{F1F1F1} 0,99}  & \multicolumn{1}{l|}{\cellcolor[HTML]{D7E8F1}{\color[HTML]{000000} -0,17}} & \cellcolor[HTML]{EDF2F5}{\color[HTML]{000000} -0,05} \\ \hline
    HadGEM3                              & \multicolumn{1}{l|}{\cellcolor[HTML]{F6F7F7}{\color[HTML]{000000} 0,00}}  & \cellcolor[HTML]{B8D8E9}{\color[HTML]{000000} -0,28} & \multicolumn{1}{l|}{\cellcolor[HTML]{E1EDF3}{\color[HTML]{000000} -0,12}} & \cellcolor[HTML]{F9EFE9}{\color[HTML]{000000} 0,06}  & \multicolumn{1}{l|}{\cellcolor[HTML]{E3EDF3}{\color[HTML]{000000} -0,10}} & \cellcolor[HTML]{F9EBE3}{\color[HTML]{000000} 0,08}  & \multicolumn{1}{l|}{\cellcolor[HTML]{9C1127}{\color[HTML]{F1F1F1} 0,85}} & \cellcolor[HTML]{900D26}{\color[HTML]{F1F1F1} 0,88}  & \multicolumn{1}{l|}{\cellcolor[HTML]{F5A886}{\color[HTML]{000000} 0,38}}  & \cellcolor[HTML]{F5AA89}{\color[HTML]{000000} 0,38}  \\ \hline
    MPIESM                               & \multicolumn{1}{l|}{\cellcolor[HTML]{DAE9F2}{\color[HTML]{000000} -0,15}} & \cellcolor[HTML]{FCDECD}{\color[HTML]{000000} 0,18}  & \multicolumn{1}{l|}{\cellcolor[HTML]{D1E5F0}{\color[HTML]{000000} -0,20}} & \cellcolor[HTML]{EF9979}{\color[HTML]{000000} 0,43}  & \multicolumn{1}{l|}{\cellcolor[HTML]{D1E5F0}{\color[HTML]{000000} -0,20}} & \cellcolor[HTML]{E37E64}{\color[HTML]{000000} 0,51}  & \multicolumn{1}{l|}{\cellcolor[HTML]{C84440}{\color[HTML]{F1F1F1} 0,67}} & \cellcolor[HTML]{F5F6F7}{\color[HTML]{000000} -0,01} & \multicolumn{1}{l|}{\cellcolor[HTML]{F9EDE5}{\color[HTML]{000000} 0,08}}  & \cellcolor[HTML]{F7F6F6}{\color[HTML]{000000} 0,01}  \\ \hline
    MRI                                  & \multicolumn{1}{l|}{\cellcolor[HTML]{4F9BC7}{\color[HTML]{000000} -0,57}} & \cellcolor[HTML]{3A87BD}{\color[HTML]{000000} -0,65} & \multicolumn{1}{l|}{\cellcolor[HTML]{59A1CA}{\color[HTML]{000000} -0,54}} & \cellcolor[HTML]{3C8ABE}{\color[HTML]{000000} -0,63} & \multicolumn{1}{l|}{\cellcolor[HTML]{569FC9}{\color[HTML]{000000} -0,55}} & \cellcolor[HTML]{3C8ABE}{\color[HTML]{000000} -0,64} & \multicolumn{1}{l|}{\cellcolor[HTML]{BE3036}{\color[HTML]{F1F1F1} 0,73}} & \cellcolor[HTML]{C84440}{\color[HTML]{F1F1F1} 0,68}  & \multicolumn{1}{l|}{\cellcolor[HTML]{E1EDF3}{\color[HTML]{000000} -0,11}} & \cellcolor[HTML]{9DCBE1}{\color[HTML]{000000} -0,37} \\ \hline
    Mean                                 & \multicolumn{1}{l|}{\cellcolor[HTML]{DDEBF2}{\color[HTML]{000000} -0,14}} & \cellcolor[HTML]{D2E6F0}{\color[HTML]{000000} -0,19} & \multicolumn{1}{l|}{\cellcolor[HTML]{E0ECF3}{\color[HTML]{000000} -0,12}} & \cellcolor[HTML]{F3F5F6}{\color[HTML]{000000} -0,02} & \multicolumn{1}{l|}{\cellcolor[HTML]{DDEBF2}{\color[HTML]{000000} -0,14}} & \cellcolor[HTML]{F0F4F6}{\color[HTML]{000000} -0,04} & \multicolumn{1}{l|}{\cellcolor[HTML]{AE172A}{\color[HTML]{F1F1F1} 0,81}} & \cellcolor[HTML]{CB4942}{\color[HTML]{F1F1F1} 0,66}  & \multicolumn{1}{l|}{\cellcolor[HTML]{FBE3D4}{\color[HTML]{000000} 0,15}}  & \cellcolor[HTML]{FAE7DC}{\color[HTML]{000000} 0,11}  \\ \hline
    \end{tabular}
        
    \end{subtable}

\end{table}

\end{document}